\documentclass[%
 reprint,
 amsmath,amssymb,
 aps,
]{revtex4-2}
\usepackage{amsmath,amssymb}
\usepackage{slashed,verbatim,graphicx}
\usepackage{graphicx}
\usepackage{placeins}
\usepackage{ytableau}
\usepackage{pstricks}
\usepackage{color}
\usepackage{tikz}
\usepackage{qcircuit}
\usepackage{physics}
\usepackage[pdftex, pdfstartview={FitH}, pdfnewwindow=true, colorlinks=false, pdfpagemode=UseNone]{hyperref}
\usepackage[caption=false]{subfig}
\usepackage{float}

\def\Tr{\textrm{Tr}}

\newcommand{\ignore}[1]{}
\newcommand\be{\begin{equation}}
\newcommand\ee{\end{equation}}
\newcommand\bea{\begin{eqnarray}}
\newcommand\eea{\end{eqnarray}}
\newcommand{\bdm}{\begin{displaymath}}
\newcommand{\edm}{\end{displaymath}}
\newcommand{\bal}{\begin{align}}
\newcommand{\eal}{\end{align}}
\newcommand\nn{\nonumber \\}
\newcommand{\<}{\langle}

\renewcommand{\>}{\rangle}
\renewcommand\dd{\partial}

\ytableausetup{aligntableaux=top, boxsize=1em}

\begin{document}

\title{ $U(1)$ Fields from Qubits: 
an Approach via D-theory Algebra}

\author{David Berenstein}
\author{Hiroki Kawai}
\affiliation{Department of Physics, University of California at Santa Barbara, CA 93106}

\author{Richard Brower}
\affiliation{Department of Physics, Boston University, Boston, MA 02215}

\date{\today}

\begin{abstract}
A new quantum link 
microstructure was proposed for the lattice quantum chromodynamics (QCD) Hamiltonian, replacing the Wilson gauge links with  a bilinear of fermionic qubits, later
generalized to D-theory
.  
This formalism provides a general framework
for building lattice field theory algorithms for quantum computing
.   
We focus mostly on the simplest case of a quantum rotor for a single compact $U(1)$ field.
We also make some progress for non-Abelian setups, making it clear that the ideas developed in the $U(1)$ case extend to other groups. 
These in turn are building blocks for $1 + 0$-dimensional ($1 + 0$-D) matrix models, $1 + 1$-D sigma models and non-Abelian gauge theories in $2+1$ and $3+1$ dimensions. 
By introducing multiple flavors 
for the $U(1)$ field, where the flavor symmetry is gauged, we can  efficiently approach the infinite-dimensional Hilbert space of the quantum $O(2)$ rotor with increasing flavors.
The emphasis of the method is on preserving the symplectic algebra exchanging fermionic qubits by sigma matrices (or hard bosons) and developing a formal strategy capable of generalization to $SU(3)$ field for lattice QCD and other non-Abelian $1 +1$-D sigma models or $3 +3$-D gauge theories.     
For $U(1)$, we discuss briefly the qubit algorithms for the study of the discrete $1+1$-D Sine-Gordon equation.
\end{abstract}

\maketitle


\section{Introduction }
\label{sec:Introduction}

Lattice field theory, particularly the Wilson's formulation of quantum chromodynamics~\cite{Brambilla:2014jmp}, now plays a central role in high energy physics being capable of {\em ab initio} precise predictions in support of the search for physics beyond the standard model (BMS). 
This is due to a firm theoretical foundation, combined with spectacular advances in algorithms on classical computers soon to approach the Exascale. 
It is generally accepted that the Wilson Euclidean (imaginary-time) lattice action lies in the basin of attraction of QCD, converging to the exact answer in infinite volume (IR) and  zero-lattice-spacing (UV) limits.

However, the standard Monte-Carlo integration is incapable of
real-time dynamics. 
One way to change this paradigm could be quantum computing. 
This requires not only the development of quantum computing technology but also the transformation of the lattice field theories to an appropriate Hamiltonian $\hat H$ expressed in terms of qubits (sigma matrix operations), as first noted by Feynman in 1982~\cite{feynman1982simulating}.
The first step to convert the lattice action to a Hamiltonian formulation is straightforward. 
For example, for QCD, by taking the time continuum limit of the transfer matrix in the Wilson's lattice QCD, one obtains the Kogut-Susskind Hamiltonian~\cite{Kogut:1974ag} operator
\begin{align}
    \hat H_{QCD} &= \frac{g^2}{2}\sum_{\<x,y\>} Tr[ E^2(x,y)] \nn
    &\quad +   \frac{1}{2g^2} \sum_{x,\mu \ne \nu} Tr[2 - U_{\mu \nu}(x) - U^\dag_{\mu\nu}(x)] \nn
    & \quad + \Psi^\dag D[U] \Psi
\label{eq:gauge}
\end{align}
where $\{\expval{x, y}\}$ is the set of all of the pairs of the nearest-neighbor lattice sites with the specified direction $x\rightarrow y$, i.e. all the directed lattice links. 
The plaquette operators $U_{\mu\nu}(x)$ are defined as 
\begin{align}
    U_{\mu\nu}(x)
    &\equiv 
    U(x, x+\hat \mu) U(x+\hat \mu, x+\hat \mu + \hat \nu)\nn 
    &\quad \times 
    U^\dagger(x+\hat \nu, x+\hat \mu + \hat \nu) 
    U^\dagger(x, x+\hat \nu)
\end{align}
with the Wilson link operators $U(x,y) \equiv \exp[ i A(x,y)] $ determined by the gauge field $A(x,y)$~\footnote{On a  regular hypercubic lattice, fields
on directed links $\<x,y\>$ with a positive spatial shift, $y = x +\hat \mu$ are often labelled by  $U_\mu(x) = U(x,x+\hat \mu)$ and the somewhat awkward backward link by 
$U^\dag_\mu(x, x -\hat \mu) = U(x+\hat \mu,x)$.
The discrete curl for the magnetic term is then the path ordered product on each square: $U_{\mu\nu}(x) = U(x,x +\mu)  U(x+\mu,x + \nu) U(x+\nu,x+\mu) U(x+\nu,x)$.}
in the adjoint of the gauge group. 
We refer to $E(x,y)$, which are conjugate to the gauge fields $A(x,y)$, as the electric field operators. 
Hence, $E^2(x,y)$ is the Casimir of the gauge group. 
The quark term is $ \Psi^\dag D[U] \Psi$.
The symplectic algebra between $E(x,y)$ and $U(x,y)$ on each link $\<x,y\>$ preserves the exact spatial gauge invariance and the Gauss' law. 
It is then anticipated, based on the Osterwalder-Schrader
positivity, that the unitary evolution operator $U(t,0) = \exp[ - i t \hat H_{QCD}]$ 
of the lattice Hilbert space also converges to the  exact quantum dynamics as the UV lattice spacing and the finite volume IR cutoff are removed.

The second step, converting the problem into qubit operators, is more difficult, at least on all proposed hardware to date.
The main difficulty comes from that the local variables on a single link, when quantized, act on an infinite-dimensional Hilbert space. 
This is the function space $L^2(G)$ on the group manifold of the local gauge group $G$.
Roughly speaking, we have a wavefunction $\psi(g)$ of the classical group variable $g \in G$, which needs to be normalizable. 
For example, for QCD, the infinite-dimensional Hilbert space of the $SU(3)$ group manifold at each link must be drastically reduced. 
On modern classical computers, this is solved by the illusion of the continuum with a mild 32- or 64-bit truncation 
of floating-point arithmetic approximation. 
On the other hand, this Hilbert space must be represented by a small number of qubits per lattice site on proposed quantum hardware with a limited number of qubits at present.  
The problem is to invent a new microstructure for a qubit Hamiltonian operator that falls into the {\bf universality class} of the Kogut-Susskind Hamiltonian.  
At least in that sense, when we take the large volume and small lattice size limit, we should recover the exact QCD for the low energy states near the vacuum.

A general framework, which is referred to as Quantum Links~\cite{Chandrasekharan:1996ih,Brower:1997ha, Beard:1997ic} or more properly its generalization called D-theory~\cite{Brower:2003vy}, has been proposed to achieve this. 
In D-theory, the $E$ and $U$ fields are replaced with the quantized $\hat E$ and $\hat U$ operators, respectively, on each link. 
These operators are represented as the bilinears of a small set of fermionic operators. 
The fermionic representation is an explicit example of what Bravyi and Kitaev~\cite{Bravyi_2002} refer to as local fermionic modes (LFM) whose algebra can be represented as products of hard boson sigma matrices. 
The basic heuristic to plausibly reach the correct universality class is: (i) to wisely choose the base lattice to satisfy a maximal set of space symmetries and (ii) to find field operators which still satisfy the basic symplectic algebra of the link operators and their conjugate electric operators~\footnote{Here we mean that we exactly preserve the local symmetry of gauge transformations on a local link,}. 
It is plausible that by preserving lattice symmetries and the symplectic structure, many simple examples can be found in the basin of attraction of continuum field theory as indeed first conjectured by Feynman in 1982~\cite{feynman1982simulating}. 
Preserving the fundamental symplectic algebra opens up a range of
qubit realizations \textit{via} D-theory for efficient quantum computing as summarized recently by Wiese in~\cite{wiese2021quantum} and in an alternative qubit construction by Liu and Chandrasekharan in~\cite{https://doi.org/10.48550/arxiv.2112.02090}. 
Here, we restrict our investigation to the simplest example of field operators on the compact $G=U(1)$ group manifold. 
Already this quantum rotor provides an interesting and non-trivial 
building block for quantum spin and gauge theories.

Of course, establishing the Hamiltonian in the desired universality class is a difficult problem. 
It generally requires both theoretical insight and numerical evidence. 
The original D-theory  paper argued it for asymptotically free chiral models in $1+1$ dimensions and gauge theories in $3+1$ dimensions. 
The universality would be valid with only a logarithmic growing layering of a single qubit in an extra dimension~\cite{Schlittgen:2000xg}.
While this is a modest increase in the volume, the discovery of
other options is anticipated by the 
evidence found in~\cite{Bhattacharya:2020gpm} of a 
lattice Hamiltonian for the $1+1$-D non-linear $O(3)$ sigma model with only two layers. 
The qubit systems exhibit both the UV asymptotic free fixed point and the IR universality in the continuum.  
For our $U(1)$ example, the reader is also referred to the study by Zhang, Meurice and Tsai~\cite{Zhang:2021dnz}.
In their work, it is noticed that the Berezinskii-Kosterlitz-Thouless (BKT) phase transition, which is expected for the continuum 2-D $O(2)$ (XY) model, is absent for 3-states truncation per site but appears for 5-states truncation or more. 
The lesson here is that if the truncation is too drastic, one might be outside of the desired universality class. 

Here, we consider the limited question of how the use of $M$ copies of fermionic qubits (referred to as a flavor index in
\cite{Schlittgen:2000xg,Bar:2001gz}) at each link can converge locally to the Kogut-Susskind Hamiltonian as $M \rightarrow \infty$.
This sequence provides a qubit implementation that can be explored with respect to universality and efficient quantum computing with the hope that very few qubits per lattice volume suffice. 
This paper is also restricted to the simplest example as we mentioned: a compact $U(1)$ field manifold formulated in a way that is capable of generalization to non-Abelian group manifolds. 
We would have the finite approximation of $L^2(S^1)$, the Hilbert space of the $U(1)$ theory we study, as the quantization of the local variable. 
Even in this Abelian example, the Lagrangian formalism is mapped to a nontrivial $SU(2)$ quantum rotor as a Hamiltonian, a basic ingredient of the qubit codes and even their hardware realization~\cite{Raynal_Kalev_Suzuki_Englert_2010, Albert:2017}. 
Applications are interesting for a variety of quantum field theories, not just for gauge theories. 
Depending on if we have certain gauge constraints or
not, what matters is the fact that the fields give an interesting local Hilbert space structure at a site or link. 
The main analysis of local fields can be applied to examples such as the XY model, the Sine-Gordon theory or the Schwinger model in $1+1$ dimensions and gauge theories in $2+1$ and $3+1$ dimensions. 
For example, in the discretized version of the Sine-Gordon model, the local variable can also be taken to be a
periodic variable living on each of the lattice sites rather than
the links. 
A similar comment would be applied to non-linear sigma models on group manifolds where we would obtain $L^2(G)$ at each site, rather than $L^2(G)$ on links with the Gauss' law constraints.
In this sense, this paper is more concerned with the individual
manifold for local fields either on a link or lattice site, rather
than the problem of a full quantum theory. 
We are basically asking how to generate local variables that become bosons (with a non-trivial manifold and symmetry structure) when the cutoff on the local variable is removed, while the symmetry structure is realized exactly.

The paper is organized as follows. 
In Sec.~\ref{sec:Algebra}, we present the general algebraic constraint of quantum links for the $U(N)$ field with multiple flavors which is specialized to $U(1)$, and we also comment on how the quantum links with gauged flavor give a description that is a truncation of the Hilbert space of more general group manifolds with no additional states.  
In Sec.~\ref{sec:Truncation}, we define 
the truncation of the $U(1)$ quantum Hamiltonian both
for the D-theory flux cutoff and the $\mathbb Z_N$ clock rotor fields truncation. 
In Sec.~\ref{sec:qubit} we present the translation of the $U(1)$ quantum link operators with fermionic operators to those with sigma matrices. 
In Sec.~\ref{sec:spectral-matching}, we numerically compare the spectra of the truncated models in our formalism as well as that of the $\mathbb Z_N$ clock rotor fields truncation. 
Sec.~\ref{sec:Applications} considers briefly
the quantum circuits to implement the $1+1$-D XY and Sine-Gordon models for the lowest triplet truncation and study the phase transition by measuring the entanglement entropy of the ground states. 
In Sec.~\ref{sec:Discussion}, we elaborate further on our results.

\section{Symplectic Algebra and Universality}
\label{sec:Algebra}
A Hamiltonian for a classical mechanical system is defined
by the symplectic structure of its {\em P-Q} coordinates expressed as the Poisson brackets. 
A quantum Hamiltonian, just as the classical case, is also defined by the symplectic structure, promoting the Poisson brackets to the canonical commutators. 
Using the Kogut-Susskind Hamiltonian as an example to motivate the D-theory construction, we first double the phase space introducing a left-right pair, $ E_L(x,y)$, $ E_R(x,y)$ electric fields or gauge generators on each link and a pair of forward and backward link
operators $U(x,y)$ and $U(y,x) = U^\dag(x,y)$.
\begin{align}
\hat H &= \frac{g^2}{4}\sum_{\<x,y\>} Tr[ E^2_L(x,y) +  E^2_R(x,y)]\nn 
&\quad + \frac{1}{2g^2}\sum_{x, \mu \ne \nu}Tr[2 - U_{\mu, \nu}(x) - U^\dag_{\mu, \nu}(x)]
\end{align}
The fermionic matter term, $\Psi^\dag D[U] \Psi$, is straightforward to be added, but not essential for our current discussion. 
At first, it might seem strange that one has to double the variables.
This is quite natural when one is studying motions on a group manifold. 
This is because we have two possible group actions on $G$, by the left and right multiplications. 
There are a set of generators for each of these transformations, i.e. the electric fields. 

The full symplectic algebra on each link $\<x,y\>$ in the doubled phase space is summarized as 
\begin{align}
    &[E^\alpha_L, U] = \lambda^\alpha U, 
\quad \quad 
[E^\alpha_L, U^\dag] =  - U^\dag \lambda^\alpha,\nn
&[E^\alpha_R, U] = - U \lambda^\alpha, 
\quad [E^\alpha_R, U^\dag ] = \lambda^\alpha  U^\dag 
\label{eq:symplectic}
\end{align}
where the $\lambda^\alpha$ matrices are the generators of $G$ in the fundamental representation. 
$E_L$ and $E_R$ generate two independent copies of $G$, namely $G_L$ and $G_R$ respectively: 
\begin{align}
    &[E^\alpha_{L}, E^\beta_{L}] = if^{\alpha \beta \gamma} E^\gamma_{L}, \quad  
[E^\alpha_{R}, E^\beta_{R}] = if^{\alpha \beta \gamma} E^\gamma_{R}, \nn 
&[E^\alpha_{L}, E^\beta_{R}] = 0
\label{eq:gauge_algebra}
\end{align}
In other words, the $U$ variables transform in the representation of $(\mathbf{fund.}, \overline{\mathbf{fund.}})$ of $G_L \times G_R$ rather than the adjoint of $G$ as in the ordinary construction of gauge theories, where $G_L$ is generated by $E_L$ and $G_R$ is generated by $E_R$, while the $U^\dagger$ variables are in $(\overline{\mathbf{fund.}}, \mathbf{fund.})$.  

It is also known that it is convenient to study the left and right invariant forms, $U^{-1}d U$ and $d U U^{-1}$, which lead to velocities $v_L= U^{-1} \dot U$ and $v_R= \dot U U^{-1}$. 
Each of these can serve as a basis for velocities, and they are clearly related to each other by 
\begin{equation}
    v_R= U v_L U^{-1}. \label{eq:constraint_vel}
\end{equation} 
When one is careful with these velocities, we get canonical conjugates to the group variables that encode the symmetry. 
These are Lie algebra valued, generating group transformations in the Hamiltonian sense, and the left and right actions on $G$ commute with each other.

The original Hamiltonian is then recovered
with the constraint of unitarity and the constraint inherited from the velocities Eq.~\eqref{eq:constraint_vel} on each link $\<x, y\>$
\begin{align}
    &U^\dag(x,y) U(x,y) = 1, \nn
    &E_R(x,y) =  U(x,y)E_L(x,y)U^\dag(x,y)  
\label{eq:constraintsU}
\end{align}

Preserving the symplectic structure would mean that either we keep Eq.~\eqref{eq:symplectic} and Eq.~\eqref{eq:gauge_algebra}, or we keep Eq.~\eqref{eq:constraintsU}.
If we keep both, we have the full $L^2(G)$ which is infinite-dimensional.

\ignore{
The advantage of the doubling of the Q-P coordinates is that
the new microstructure of D-theory   sacrifices only the exact unitary constraint $U
U^\dag = 1$, which is an inessential choice of
the UV cutoff. For example in gauge theories, the
renormalization from smaller scale for color transport, is
summed over different unitary ordered  paths not obeying this
constraint.}

\subsection{Fermionic D-Theory Algebra}
\label{sec:Dtheory}
In this section, we specifically pick the gauge group to be $G = U(N)$. 
This still demonstrates the more general framework of the D-theory discretization than that for the more simpler Abelian case, $U(1)$, which will be our main focus for the later sections.

A straightforward  discrete representation that exactly preserves the symplectic algebra in Eqs.~\eqref{eq:symplectic}-\eqref{eq:gauge_algebra} replaces the single link field on a compact group manifold by a bilinear of fermion operators as
\begin{align}
    &U^i_j \rightarrow \hat U^i_j = \vec a^i \cdot \vec b^\dag_j  =  \sum_{m=1}^M (a_m b^{\dag m})^i_j, \nn 
  &(U^\dag)^j_i \rightarrow  (\hat U^\dag) ^j_i= \vec b^j \cdot \vec a^\dag_i = \sum_{m=1}^M (b_m a^{\dag m})^j_i
\label{eq:ab}
\end{align}
Notice that the matrix elements of the link operators are no longer complex numbers, but rather operators. 
We denote that by putting the ``hat" notations on top of the operators. 
The scalar product implies a sum over the vector of $M$ flavors of creation and destruction operators: $\vec a^i = (a^i_1,a^i_2 \cdots a^i_M)$, $\vec b^j = (b^j_1,b^j_2, \cdots, b^j_M)$.
The indices $i$ and $j$ are color indices running from 1 to $N$. 
All the $4 N M$ fermionic
operators $a^i_m(x,y)$ and $b^i_m(x,y)$ per link
obey the standard anti-commutator relations of single fermionic degrees of freedom, as introduced
in \cite{Schlittgen:2000xg,Bar:2001gz}.  
The symplectic algebra Eq.~\eqref{eq:symplectic} fixes the representation of the electric flux:
\begin{align}
    &(\hat E_L)^i_j =   \vec a^\dag_j \cdot\vec a^i =  \sum_{m=1}^M (a^{\dag m} a_m)^i_j, \nn 
    &(\hat E_R)^i_j  = \vec b^\dag_j \cdot\vec b^i =  \sum_{m=1}^M (b^{\dag m} b_m)^i_j 
\end{align}
reproducing the exact gauge algebra in Eq.~\eqref{eq:gauge_algebra}.
Although this seems cumbersome, the $a$ operators carry the left
action and the $b$ operators carry the right action. 
In this way, $E_L$ and $E_R$ have been separated into completely distinct variables. 
Each flavor of $a$ carries the same representation with respect to the left
Lie algebra: the fundamental. 
The same is true for $b$, but carrying the antifundamental.
The flavor index $m$ only appears in sums, so the flavor symmetry $U(M)$ can be thought of as a local constraint on each link. 
This constraint ties the left and the right actions to each other eventually.

The resulting fermionic qubit form, referred to in Bravyi and
Kitaev~\cite{Bravyi_2002} as \textit{ local fermionic modes}, is
a small finite Fock space on each lattice link. 
The original link variables $U(x,y)$ commute with each other resulting from the unitarity constraint Eq.~\eqref{eq:constraintsU}, whereas in the fermionic representation, this is not maintained. 
The only non-zero commutator is local to each link: 
\begin{align}
    &[\hat U^i_j(x,y),  (\hat U^\dag)^k_l (x',y')] = \delta_{xx'}\delta_{yy'}((\hat E_L)^i_l \delta^k_j - (\hat E_R)^k_j \delta^i_l)\nn 
    & \implies  [ \hat U(x,y), \hat U^\dag(x',y') ] ^i_j= \delta_{xx'}\delta_{yy'} N ((\hat E_L)^i_j - (\hat E_R)^i_j)
\end{align}
Thus, a link matrix is no longer normal and as a consequence, breaks the unitarity constraint.  
The symplectic algebra at each link treats $E_L$ and $E_R$ as  independent {\em velocity coordinates}, conjugate to  non-commuting {\em position} operators, $\hat U$ and $\hat U^\dagger$.

This breaking should be interpreted in its entirety as an irrelevant UV cutoff effect. 
As we go to the continuum limit, with sums of
multiple paths between distant sources, this non-commutation due to the infrequent intersection at the cutoff scale should vanish. 
Moreover, when averaging over paths for long
distances, we would also abandon the constraint $U^\dagger U=1$, which
is not satisfied when we use the expectation values for $U$ and
$U^\dagger$ separately.
 
It should also be noted that this construction of operators in the multi-flavor fermionic Hilbert space satisfying the symplectic algebra is not unique.
Rather, it provides a general framework with multiple solutions which can be adapted to better approximate the infinite-dimensional Hilbert space with finite-dimensional space and provide alternative qubit implementations to optimize quantum codes. 
Indeed, this flexibility of the D-theory framework is what we exploit in the current application for $U(1)$.  
As we will show explicitly for the $U(1)$ example, the multi-flavor fermionic space factorizes into super selection sectors which can be modified to give a sequence of bosonic qubit models restoring the
zero commutator in the limit of $M \rightarrow \infty$.
It is useful to construct a variety of D-theory candidates to
explore our U(1) examples, which will be carried in Sec.~\ref{sec:qubit}. 

It is important to note that the discrete representation of the link and the electric operators constructed above are the generators of the $U(2N)$ Lie algebra: 
\begin{align}
\begin{bmatrix}
\; \hat E_L  &\; \; \hat U^\dag \;  \\
\; \hat U  &\; \; \hat E^R \; 
\end{bmatrix}^i_j =
\begin{bmatrix}
\;\vec a^\dag \cdot \vec a& \; \;  \vec a^\dag \cdot \vec b \;\\
 \; \vec b^\dag \cdot \vec a  &\; \;  \vec b^\dag \cdot \vec b\; 
\end{bmatrix}^i_j
\end{align} 
Hamiltonian evolution remains on this $U(2N)$ group manifold at each link. 
It is remarkable for gauge theories that the quantum link Hamiltonian preserves exactly the local symmetry rotation at each site.  
The construction of the formalism would also apply to a model with a global Lie group symmetry with a compact manifold. 
One example is the spin models such as the $1+1$-D chiral theory
\begin{align}
    \hat H_{chiral} &= \sum_x Tr[\hat E^2_L(x)] + Tr[\hat E^2_R(x)] \nn 
    & +\lambda \sum_{\<x,y\>}Tr[2 -\hat U(x) \hat U^\dag(y) - \hat U(y) \hat U^\dag(x)]
\end{align}
with global $U(N)$ symmetries. 
The term with the coupling $\lambda$ is the square of the discretized differentiation $\left(\hat U(x)-\hat U(y)\right)^\dagger \left(\hat U(x)-\hat U(y)\right)$.
The spin theory will have global symmetry generators $\hat J_L = \sum_x \hat E_L(x)$ and $\hat J_R  = \sum_R  \hat E_R(x) $ so that $[\hat J_L, \hat H]  =
[\hat J_R, \hat H] = 0$, where all fields transform as $\hat U(x) \to g \hat U(x)h^{-1}$ for common $g,h$.
Precisely determining whether or not this radical reduction of the degrees of freedom is still capable of reaching a universal continuum fixed point is generally a difficult dynamical question. 
We will not attempt to solve this problem here.

We also refer the reader to the reference~\cite{Brower:2003vy} for other group manifolds. 
For example, the algebraic structure for $SO(N)$, $SU(N)$, and $Sp(N)$ gauge theories naturally lie in $SO(2N)$, $SU(2N)$ and $Sp(2N)$ algebras respectively and as well as the $O(N)$, $U(N) \otimes U(N)$ quantum spin models.

\subsection{Restoration of the continuum Hilbert space\label{sec:Peter-Weyl}}


Our goal here is to show that when $M\to \infty$ we should recover
the Hilbert space of the original $U$ variables that would enter in
the Kogut-Susskind formulation. 
Although this illuminates our method, the reader may choose to go directly to more intuitive geometrical interpretation discussed in Sec.~\ref{sec:geometry} or the concrete construction carried out in Sec.~\ref{sec:Truncation} on the $U(1)$ example.  
It is possible to show in general that the state space
is easily projected into a subspace with each link represented
by a few hard boson (or sigma matrix) degrees of freedom. 
This representation is trivial for $U(1)$ and only requires a local
Jordan-Wigner transformation inside the group at each link.
In that formulation, the entries of the matrices $U$ are
scalar functions of the group elements. 
These also commute with each other and their polynomials generate the space of $L^2$ functions on the group manifold $G$. 
The Hilbert space $L^2(G)$ itself is given by the following definition.
We need wavefunctions from the group manifold to the complex numbers
\begin{equation}
    \psi: G\to {\mathbb C}
\end{equation}
with the inner product implemented as 
\begin{equation}
    \braket{\psi}{\varphi}= \int dg\; \psi^*(g)\varphi(g) 
\end{equation}
where $dg$ is the Haar measure on the group manifold, which is the unique group invariant measure.
The trivial function $\psi(g) =1$ is group invariant. 
All other wave functions can be obtained from this by polynomials of the $U,U^{-1}$ matrix component functions and then taking 
the $L^2$ completion. 

We want to show that our quantum link procedure approximates this $L^2(G)$ Hilbert space. 
It is convenient for us to consider a slightly modified realization of
the $U$ variables as bilinear of the fermions.  
As described in Eq.~\eqref{eq:ab}, the operators $\hat U$ and $\hat U^\dagger$ leave a
total occupation number unchanged. 
There is an automorphism of fermion
algebras $ a^j_m\leftrightarrow c^{\dagger j}_m$ and
$a^{\dagger m}_j\leftrightarrow c^m_j$ which makes it possible to
describe $\hat U$ as made purely from raising operators and $\hat U^\dagger$ from lowering operators. 
Namely, they become $\hat U\propto a\cdot b^\dagger \equiv c^\dagger\cdot b^\dagger $ and
$\hat U^\dagger\propto a^\dagger \cdot b \equiv c\cdot b$. 
As we noted, the contractions of the flavor indices can be thought of as gauging the $U(M)$ symmetry. 
If we also include the left Lie algebra action of $U(N)$ and the right action, the degrees of freedom on a link are charged
under the $U(N)_L\times U(M)\times U(N)_R$ symmetry.  
Under this symmetry, the operators transform as Table~\ref{tab:fermion-reps}. 
\begin{table}[t]
    \centering
    \begin{tabular}{c|c c c}
         &  $U(N)_L$ & $U(M)$ & $U(N)_R$ \\
         \hline \hline 
         $c^\dagger$ & $\mathbf{N}$ & $\overline{\mathbf{M}}$ & $\mathbf{1}$\\
         $b^\dagger$ & $\mathbf{1}$ & ${\mathbf{M}}$ & $\overline{\mathbf{N}}$\\
         \hline 
         $\hat U \propto c^\dagger\cdot b^\dagger$ & $\mathbf{N}$ & $\mathbf{1}$ & $\overline{\mathbf{N}}$\\
    \end{tabular}
    \caption{The representations of the fermion operators as well as the bilinear $\hat U$ under the left and right color gauge symmetries and the flavor gauge symmetry. The conjugate annihilation operators are in their conjugate representations. }
    \label{tab:fermion-reps}
\end{table}
The advantage of this setup is that the standard vacuum of the $b,c$ fermions is neutral with respect to all the symmetries, hence it is gauge invariant. 
Let us call this standard vacuum $\ket{\Omega}$. 
We can reach other gauge invariant states by acting on $\ket\Omega$ with gauge invariant operators under the $U(M)$, namely, the matrix elements of $\hat U$ and $\hat U^\dagger$.
Notice that $\hat U^\dagger\ket{\Omega}=0$ but $\hat U$ does not annihilate it. 
This means that $\hat U$ and $\hat U^\dagger$ act asymmetrically on the reference state $\ket{\Omega}$. 
The complete set of states is built by acting with many $\hat U$ operators.
The $\hat U$ operators commute with each other, so their actions are just as commuting bosonic generators.

The Hilbert space obtained this way can be decomposed into the irreducible representations of $U(N)_L \times U(N)_R$. 
A state $\hat U^n\ket{\Omega}$ has $n$ upper indices with respect to $U(N)_R$ and $n$ lower indices with respect to the left $U(N)_L$. 
Because of the bosonic statistics, permutations of upper indices can be undone by a change in the order of the product, so long as the permutation is turned over to the lower indices. 
Projecting into different representations is done by these permutations, and it corresponds to a Young diagram (tableau) representation with $n$ boxes. 
One of the diagrams for e.g. $n = 10$ is 
\begin{align}
    Y(5,3,2)= \ydiagram[]{5,3,2}
\end{align}
The diagram for the lower indices is the same, but since the indices are lowered, they are in the conjugate representation. 
In the intermediate flavor index, the fermionic statistics requires transposing the Young diagram. 
This argument appeared in \cite{Berenstein:2004hw} (see also \cite{Berenstein:2019esh} and references therein).
The Hilbert space can be therefore decomposed into the sum of the tensor products of an irreducible representation of $U(N)$ and its conjugate, where each representation $R_Y$ is classified by a Young diagram $Y$:
\begin{equation}\label{eq:PW-approx}
     Hilb\sim
    \bigoplus_Y\left(\overline R_Y\otimes R_Y\right)\sim \bigoplus_Y Hilb(Y)
\end{equation}
Here in the Hilbert space, each summand is represented by one copy of the Young diagram for the upper indices let's say, with the understanding that the conjugate representation is giving the representation of the other $U(N)$ in the lower index structure.

We need to show that when we take
$M\to\infty$ of this Hilbert space Eq.~\eqref{eq:PW-approx} in our quantum link formulation, we can recover the Hilbert space $L^2(U(N))$ for the Kogut-Susskind formulation. 
The constant function $\psi(g) = 1 \in L^2(U(N))$ plays the role of the vacuum $\ket 0$. 
The excited states are described by the harmonic functions on $U(N)$.
In $L^2(U(N))$, both $U$ and $U^\dagger$ act 
non-trivially on the vacuum, whose actions are different from the actions we have on the fermion reference vacuum state $\ket{\Omega}$. 
This demands us to find the correct vacuum state $\ket 0$ in the Hilbert space of $\hat U$ corresponding to $\psi(g) = 1$. 
For any Lie group $G$, we can now appeal to the Peter-Weyl theorem.
This theorem states that when we decompose $L^2(G)$ into representations of the left ($G_L$) and right ($G_R$) symmetries of the group multiplication, we get that $G_L\times G_R$ is decomposed into a direct sum of the products of their irreducible representations
\begin{equation}
 L^2(G) = \bigoplus_R \overline R \otimes R    .
\end{equation}
In this sum, all irreducibles of $G$ appear exactly once. 
If we compare it to the description above around $\ket{\Omega}$, we obviously have a mismatch: the $U(N)$ representations are classified by pairs of Young diagrams with some constraints rather than with a single Young diagram. 
In the double Young diagram, one Young tableaux is for boxes (they count powers of $U$) and the other one is for antiboxes
(they count powers of $U^\dagger$) \cite{Gross:1993hu}. 
The constraint is that the longest column of the box tableau plus the longest column of the antibox tableau need to add up to less than or equal to the rank of the group, in this case $N$. This is the constraint that says contractions are trivial  $U^a_b (U^\dagger)^b_c =\delta^a_b$. 

Let us look at how one of these pairs of tableaux, denoting a single representation, can be represented graphically. For example, for $U(5)$, we can take
\begin{equation}
\ydiagram[]{3,2}\times \ydiagram[\bullet]{3,1\quad} \equiv
\ytableausetup{notabloids}
  \begin{ytableau}
\none&   \none & \none & \  & \ & \ \\
 \none&  \none &   \none   &\  &\  \\
  \none&  \none&\none \\
\none& \none & \bullet \\
    \bullet &  \bullet &\bullet
  \end{ytableau}\label{eq:doublediag}
  \end{equation}
The second tableaux with the filled boxes is the one with anti-boxes.
It is turned 180 degrees and put at the bottom of the diagram. The total vertical size is $N$ ($=5$ in this case). 
The constraint is such that the two tableaux do not overlap horizontally.

The main idea to show that we can write the Hilbert space with these pairs of Young diagrams in terms of single Young diagrams is as follows. 
We choose as a reference state a tableau that is filled all the way down to the bottom $N$ rows, with $K$ boxes 
on each row. 
That is, we choose as a new vacuum a tableau that is actually a singlet of $SU(N)$, but that carries $U(1)$ charge $NK$. That is, we choose as a new vacuum a Young diagram (for $N = 5$ and $K = 6$ let us say)  
\begin{equation}
\ket 0 \equiv \ydiagram[]{6,6,6,6,6} 
\end{equation}
where we have filled the boxes up to the maximum allowed  depth $N$.
These states are unique because they are one-dimensional representations of $U(N)$, once we fix the charge.
If we want to represent Eq.~\eqref{eq:doublediag} relative to this ground state, we add to the reference state the boxes of the fundamentals in the upper corner and we subtract the antiboxes on the bottom corner. 
For the above example with $N= 5$, 
\begin{equation}
\ket 0 \times  \ydiagram[]{3,2}\times \ydiagram[\bullet]{3,1\quad} =\ydiagram[]{9,8,6,5,3} 
\end{equation}
Notice also that the representations of $SU(N)$ that appear on both the $L^2(G)$ and the fermion representation have the same dimension. To get the $U(1)$ charge correctly for $L^2(U(N))$ in the fermion formulation, what  we have done in practice is that we  shifted the $U(1)$ charge so that the new vacuum has trivial charge. Happily, we see that we can match the representations of $U(N)$ with a few boxes. These are the representations with small Casimir. The constraint on $M$ tells us that the maximum width of the fermion tableaux is $M$ so that to recover the Hilbert space of $L^2(U(N))$ we need to take $M\to \infty$ and 
shift the charge enough so that the room on the left to remove boxes is as large as needed. The most symmetric way to do this is to choose $K=M/2$.

This shows that at least around the new ground state $\ket 0$, we recover the representation of the Hilbert space we want, namely $L^2(U(N))$ with a cutoff that depends on $M$. 
The gauge invariance relative to the flavor $U(M)$ shows we have no additional states to worry about. 
Computing the matrix elements of $U$, $U^\dagger$ between states is beyond the scope of the present work and will be taken in more detail in a future publication.

\subsection{Geometrical Interpretation \label{sec:geometry}}
\begin{figure}[t]
    \centering
    \begin{tikzpicture}[baseline={(current bounding box.center)}]
        \draw[] (0, 1) node[anchor=south] {\vdots}; 
        \draw[] (0, -1) node[anchor=north] {\vdots}; 
        \draw[] (-0.5, -1) -- (-0.5, 1); 
        \draw[] (0.5, -1) -- (0.5, 1); 
        \draw[dashed] (0, 0) ellipse (0.5 and 0.1);
    \end{tikzpicture}
    \caption{The phase space of the original $U(1)$ theory, which has infinite volume.}
    \label{fig:U1-phase-space}
\end{figure}
\begin{figure}[t]
    \centering
    \subfloat[]{
    \begin{tikzpicture}[baseline={(current bounding box.center)}]
        \draw[dashed] (-0.5, -1) -- (-0.5, 1); 
        \draw[dashed] (0.5, -1) -- (0.5, 1); 
        \draw[dashed] (0, 0) ellipse (0.5 and 0.1);
        \draw[] (0, 0) circle (0.5);
    \end{tikzpicture}
    \label{fig:SU2-small}
    }
    \qquad \qquad 
    \subfloat[]{
    \begin{tikzpicture}[baseline={(current bounding box.center)}]
        \draw[dashed] (-0.5, -1) -- (-0.5, 1); 
        \draw[dashed] (0.5, -1) -- (0.5, 1); 
        \draw[dashed] (0, 0) ellipse (0.5 and 0.1);
        \draw[] (0, 0) ellipse (0.5 and 1);
    \end{tikzpicture}
    \label{fig:SU2-large}
    }
    \caption{(a) The approximation with a two-sphere $\sim SU(2)/U(1)$. The equater is where the low energy physics resides and want to be matched with the dynamics on the cylinder. 
    (b) Having a larger dimensional representation of $SU(2)$ (i.e. adding more flavors) corresponds to having an elongated shape with a larger volume. }
    \label{fig:SU2-phase-space}
\end{figure}
Here we present the geometrical interpretation of the above approximation of the $L^2(U(N))$ space. 
Let us consider for the time being the simplest case of $U(1)$. 
By means of the above construction with bifermions, we get the Lie algebra of $U(2)$. 
The diagonal $U(1) \subset U(2)$ plays no role, as it commutes with all the generators and therefore decouples. 
More precisely, acting with any of the other elements of the algebra will not change the $U(1)$ diagonal charge, so it will act as a c-number when we think of a physical realization. 
We are left over with the symplectic structure with the structure of the Lie algebra of $SU(2)$. 

Is there another way to motivate this? 
The answer is yes. 
The idea is that the classical phase space of the original problem of the $U(1)$ theory leads to a cylinder: the tangent bundle of the circle as Fig.~\ref{fig:U1-phase-space}. 
This has an infinite volume, and therefore the Hilbert space is infinite-dimensional. 
We can ask if there is any other two-dimensional manifold with a finite volume and a $U(1)$ symmetry. 
The answer is, not surprisingly, yes;
the two-sphere (Fig.~\ref{fig:SU2-phase-space}) satisfies that condition \footnote{Also the two-torus.}. 
The symplectic structure of the topological two-sphere can also be written in terms of the commutation relations of the angular momentum operators. 
They play the role of $x,y,z$ coordinates, but quantized. 
This would lead us to recover the formulation above in terms of $SU(2)$ without ever mentioning the fermions. 
Upon quantization, we should get a fixed $SU(2)$ representation: a fixed  value of the quadratic Casimir, corresponding to $x^2+y^2+z^2$ for the classical manifold. 
Adding more flavors in the earlier discussion with bifermions corresponds to having a larger dimension of the $SU(2)$ representation, i.e. a larger value of the quadratic Casimir leading to a larger volume. 
This phase space is the homogeneous space manifold $SU(2)/U(1)\simeq \mathbb{CP}^1$, which is the complex projective plane of dimension one. 
To quantize this quotient space, we only need to choose the magnetic flux through the sphere (we need to choose a line bundle over the projective manifold to define the allowed wave functions). 

What should be remembered is that the metric of the two-sphere does not mean much as far as the symplectic structure is concerned.
So, we could just as well have an elongated sphere. 
This is so because we are studying Hamiltonian physics on the sphere and not a sigma model. 
What matters is how different functions on the geometry generate dynamical flows. 
When we elongate the sphere further we can produce a cylinder in the limit of infinite elongation. 
We can approach the infinite volume of the cylinder this way as we desired. 
A classical Hamiltonian function on the center band of the cylinder and the one on the center band of the elongated sphere could be very similar. 
The former is usually represented by the kinetic term $p_\theta^2$, plus any small perturbation in the angular variable (the base coordinate) $\theta$. 
In the latter case, $p_\theta^2$ is replaced by $L_z^2$, where $L_z$ is one of the three angular momentum generators. 
This band is where the low energy physics of the small kinetic term is concentrated. 
At least semi-classically, one can argue that if one low energy band of the cylinder leads to the correct universality class of some favored physics, and so does it of a sphere enough elongated to have a big enough volume to capture this band.

The generalization for multi-flavor convergence to continuum non-Abelian group manifolds are more involved. 
We outlined the method based on a more group theoretical convergence to the continuum Hilbert space by means of the Peter-Weyl
theorem in Sec.~\ref{sec:Peter-Weyl}. 
One can ask how to interpret this procedure geometrically as well as the elongated sphere for the $U(1)$ case: what is the manifold to be quantized?
The structure of coherent states in \cite{Schlittgen:2000xg} seems to have the answer: for $U(N)$, it is the complex Grassmannian ${\mathbb G}(N,2N)\sim U(2N)/ U(N)\times U(N)$. 
This is also a complex geometry of dimension $2 N^2$ and thus can be viewed as a phase space. 
More importantly, it has a group action by $U(N)\times U(N)$ acting on the left, so it is a candidate phase space with the correct group action. 
At the level of Lie algebras, the Lie algebra of $U(2N)$ provides the equivalent coordinates to $x,y,z$ above. 
One can assume that this type of Grassmannian structures will be important in all such realizations for different compact groups.

\section{$U(1)$ Quantum Rotor with UV cutoff}
\label{sec:Truncation}
In this section, we demonstrate the UV cutoff that the D-theory sets for the first step towards having a finite dimensional Hilbert space for the Abelian $U(1)$ group manifold. 
In order to test the fidelity of our $U(1)$ qubit representations, we compare it to the full $U(1)$ quantum rotor
\begin{align}\label{eq:H-KS}
    H = \frac{g^2}{2} E^2 + \frac{1}{2g^2} \left(2 - U - U^\dagger\right)
\end{align}
with symplectic algebra 
\begin{align}\label{eq:E-U-algebra}
    [E,  U] =  U, \quad 
    [E,  U^\dagger] = -U^\dagger
\end{align}
with $[U,  U^\dagger] = 0$ given from the unitarity $U U^\dagger = 1$. 
The operators with this required algebra can be represented with a scalar field $\theta \in [0,2\pi)$ as $E = - i \dd_\theta$ and $U = \exp(i\theta)$. 
It is convenient to rescale the Hamiltonian in this representation by $ 1/g^2 = \sqrt{h}$ so that 
\begin{align}
    H = -\frac{1}{2} \partial_\theta^2 + \frac{h}{2} ( 2 - 2\cos \theta) \label{eq:KShtetarep}
\end{align}
The flux representation is of course just the Fourier transforms, $\braket{\ell}{\theta} = \exp(i \ell \theta)$, with the delta function normalized states $U \ket \theta = \exp(i\theta) \ket\theta$, or in the flux
basis, $E \ket \ell =  \ell \ket \ell$.
Given that $\theta$ takes a value in the compact space of $S^1$, the flux $\ell$ takes quantized values $\ell\in \mathbb Z$.  
Explicitly writing the matrix representation of the Hamiltonian in this flux basis, 
\be
    \mel{\ell'}{H}{\ell} = \frac{\ell^2}{2} \delta_{\ell', \ell} + \frac{h}{2} (2 \delta_{\ell', \ell} - \delta_{\ell'+ 1 , \ell} - \delta_{\ell' -1, \ell})
    \label{eq:Qlink}
\ee
This can be truncated by a cutoff either in the flux basis $\ket \ell$ or in the field basis $\ket \theta$. 
We will subsequently show that the multi-flavor D-theory construction can be reformulated to exactly reproduce the flux cutoff $L$ of this rotor with $M = 2 L$ fermion flavors and therefore converge exactly to the full rotor in the $M \rightarrow \infty$ limit.
The flux truncation of the infinite-dimensional Hilbert space is carried out by restricting the flux to $\ell \in [-L,L]$ \footnote{The Casimir of the charge state is $\ell^2$, so this is also a cutoff on the maximum value of the Casimir. }. 
This UV cutoff is the first step that D-theory takes; we will call this the flux truncation and the D-theory truncation interchangeably. 
To illustrate, let us write down the matrices of the operators for the $L = 2$ cutoff case: 
\begin{align}
    U &\rightarrow U_{L=2} =
\begin{bmatrix}
0 & 1 & 0 & 0 & 0\\
0 & 0 & 1 & 0 & 0\\
0 & 0 & 0 & 1 & 0\\
0 & 0 & 0 & 0 & 1\\
0 & 0 & 0 & 0 & 0
\end{bmatrix}
, \\
 E&\rightarrow E_{L=2} =
\begin{bmatrix}
-2 & 0 & 0 & 0 & 0\\
0 & -1 & 0 & 0 & 0\\
0 & 0 & 0 & 0 & 0\\
0 & 0 & 0 & 1 & 0\\
0 & 0 & 0 & 0 & 2
\end{bmatrix} \label{eq:fluxt}
\end{align}
The field truncation is more subtle. 
One can, in comparison, think of discretizing the field to the $\mathbb Z_{2L+1}$ values with $\theta = 2\pi k/(2L+1)$ with $k \in 0,1,\cdots 2L$ and choose again to restrict the flux $\ell \in [-L,L] $ with a cyclic generator $E$ for the $\mathbb Z_{2L+1} \subset U(1)$ subgroup.
This discretization gives the same dimension of the Hilbert space as the flux truncation with the same $L$. 
Illustrating the $2L+1=5$ state truncation, the operators are
\begin{align}
    U &\rightarrow U_{\mathbb Z_5} =
\begin{bmatrix}
0 & 1 & 0 & 0 & 0\\
0 & 0 & 1 & 0 & 0\\
0 & 0 & 0 & 1 & 0\\
0 & 0 & 0 & 0 & 1\\
1 & 0 & 0 & 0 & 0
\end{bmatrix}
, \\ 
 E&\rightarrow  E_{\mathbb Z_5} =
\begin{bmatrix}
-2 & 0 & 0 & 0 & 0\\
0 & -1 & 0 & 0 & 0\\
0 & 0 & 0 & 0 & 0\\
0 & 0 & 0 & 1 & 0\\
0 & 0 & 0 & 0 & 2
\end{bmatrix} 
\end{align}
We refer to this discretization as the clock model truncation. 
Note that the electric fields are identical but the
$U$ operators are different between these two approaches. 
The first flux truncation approach, i.e. the approach that D-theory takes, preserves the symplectic algebra $[E,U] =U$ and $[E,U^\dag] =- U^\dag$
but breaks the unitarity constraint $U^\dag U = 1$, whereas the clock model does the opposite, preserving $U^\dag U = 1$ but not
the symplectic algebra. 
Preserving both leads to an infinite-dimensional Hilbert space,
which is exactly what we need to avoid \footnote{As a side note, the clock model is the quantization one gets by using a two-torus cylinder geometry, rather than the sphere. The algebra is then very similar to that of a fuzzy torus, but with this choice of $E$ there is a discontinuity of the classical variable at the gluing of the upper and lower part of the cylinder that avoids a doubling of low energy degrees of freedom.}. 
More specifically, the flux truncation (i.e. D-theory) the unitarity is violated 
with the non-zero commutator
\be
[U,  U^\dag]_{\ell',\ell} =  \delta_{\ell',L} \delta_{\ell,L}   - \delta_{\ell',-L} \delta_{\ell,-L}
\ee
Notice that these are concentrated on the largest $\ell$ exclusively, so they can be considered as only living in the UV region of the model, keeping the infrared physics roughly the same.

\begin{figure}[t]
    \centering
    \subfloat[]{\label{fig:UplusUdag3}
    \includegraphics[width=.9\linewidth]{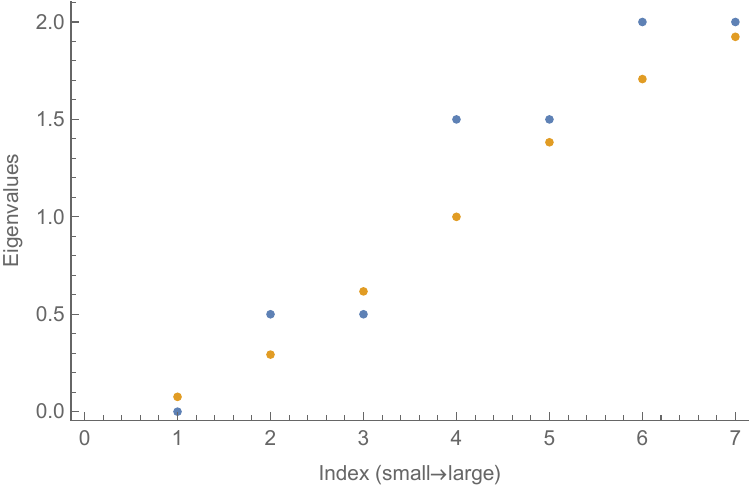}
    }\\
    \subfloat[]{\label{fig:UplusUdag8}
    \includegraphics[width=.9\linewidth]{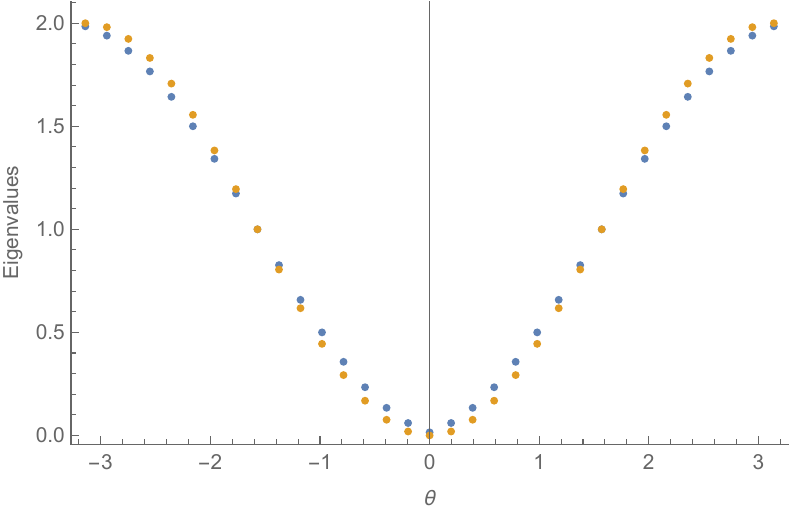}
    }
 \caption{(a)~The $7$ eigenvalues for the clock model
 potential $V(\theta) =  1 - \cos( \theta)$ (blue dots) are compared
 to  the  eigenvalues of the D-theory potential $ V = 1 - ( U + U^\dagger)/2$ (orange dots). 
 (b)~The clock model eigenvalues $V(\pi m/L)$ (blue dots) are mapped to $\theta \in [-\pi,\pi]$ compared with the  D-theory spectra (orange dots)  for $L=8$ duplicated by reflection: $\theta \rightarrow -\theta$.}
    \label{fig:UplusUdag}
\end{figure}

In Sec.~\ref{sec:spectral-matching}, we compare the low spectrum as a function of $h$ in the strong
coupling limit $h = 0$ and the weak coupling limit $h = \infty$. 
We show that the low spectra are of course exact at  $h = 0$ and remarkably accurate for a large range of values of  $h = 1/g^4$ even for a $L = 2$ or $L = 4$ flux cutoff. 
This appears to be remarkable or even paradoxical since  
for the flux truncation, the field variables obey the nilpotency $U^{2L +1} = 0$ and therefore have exactly degenerate zero eigenvalues. 
This would seem to be a poor starting point in comparison with the eigenvalues of the clock model $e^{ 2\pi k/(2L + 1)}$. 
In the clock model truncation, $U$ and $U^\dagger$ are normal, so both their real and imaginary parts are Hermitian matrices that commute and can be diagonalized simultaneously. 
Hence, their eigenvalues can be measured simultaneously, and we can use that double measurement to determine the phases $e^{ 2\pi k/(2L + 1)}$. 
For the flux truncation, on the other hand, the point is that $U$ itself does not quite have a direct correspondence to the quantum rotor field $U = \exp(i\theta)$. 
The physical correspondence becomes legit for the flux truncation once we take the combinations such as $U+U^\dagger$ which now becomes Hermitian as illustrated in Fig.~\ref{fig:UplusUdag}. 
On Fig.~\ref{fig:UplusUdag3}, we observe a nearly harmonic oscillator low spectra (orange) for the D-theory truncation even for $L=3$. 
Fig.~\ref{fig:UplusUdag8} shows a remarkable match for D-theory for all angles even at a small cutoff $L=8$. 

Note also that the field truncation with only discrete symmetry surviving has no natural generalization to non-Abelian groups; there is no infinite sequence of finite 
discrete subgroups that uniformly populate their manifolds. 
For example for $SU(2)$, the largest such finite group that is uniform on $SU(2)$ is the $120$-element icosahedral group,
and it is known that it fails to be in the universality class of the two-color gauge theory~\cite{CREUTZ1983201}.

\section{Complexity of qubit Realizations}
\label{sec:qubit}

We now turn to our multi-flavor framework with D-theory rotor
Hamiltonian Eq.~\eqref{eq:H-KS} replacing the variables $U$ and $U^\dagger$ with the operators $\hat U$ and $\hat U^\dagger$ given by a sum over the $M$-flavor fermions: 
\begin{align}
    \hat U^\dag  = \sum_{m = 1}^M a^\dag_m b^{m}, \quad 
    \hat U = \sum_{m = 1}^M b^\dag_m a^{ m}
    \label{eq:sums}
\end{align}
There are $2M$ fermions in total. 
Remembering the Fock space for a single fermion is two-dimensional, either $\ket 0$ (unfilled) or $\ket 1$ (filled), the total Hilbert space on which $\hat U$ and $\hat U^\dagger$ are acting has the dimension $4^M$.   
We note that the fermionic operators imply the nilpotency $\hat U^{M+1} = 0$ which coincides with the flux truncation with $M = 2 L$. 
However, we will see that it does not represent the same matrices for this truncation.

The $M$-flavor fermionic D-theory form starts in a $4^M$-dimensional Hilbert space, but if we impose the half-filling condition for each flavor due to the fermion number conservation as 
\begin{align}
     a^{\dagger m} a_m + b^{\dagger}_ m b^m = 1 \text{ (No sum) }
\end{align}
we are now in the Hilbert space of dimension $2^M$ allowing us to represent the operators as $M$ qubits or hard bosons.
The point to be made is that in the sums Eq.~\eqref{eq:sums}, only terms that preserve each of all of these individual fermion number combinations of $a+b$ appear, so they can be diagonalized ahead of computations.
These actually generate a subgroup of the original $U(M)$ flavor symmetries which are flavor diagonal (that is, this is a set of $U(1)^M$ generators that has been fixed). 

In this subspace, for each
flavor, we have the isometric  mapping
\be
\{a^\dag b, b^\dag a, a^\dag a - b^\dag b \} \rightarrow \{\sigma^+, \sigma^-, \sigma^3\}\label{eq:gaugeinv}
\ee
and there are no fermion statistics in the $\sigma$ on the right.
The sigma matrices are given as 
\begin{align}
    \sigma^+ = \begin{bmatrix}
        0 & 1 \\ 0 & 0
    \end{bmatrix}, \; 
    \sigma^- = \begin{bmatrix}
        0 & 0 \\ 1 & 0
    \end{bmatrix}, \; 
    \sigma^3 = \begin{bmatrix}
        1 & 0 \\ 0 & -1
    \end{bmatrix}
\end{align}
We have in this way eliminated the need to do any brute force Jordan-Wigner transformations to convert fermions into qubits. 
We identify the fermionic basis states with the spin states as the $a$ fermion filled state as $\ket \uparrow$ and the $b$ filled state as $\ket \downarrow$. 
In this form, the full representation of $SU(2)$ available by these $M$ flavors is $\bigotimes_{m=1}^M \mathbf{2}$. 
This representation is reducible, and its irreducible decomposition contains the irreducible $\mathbf{M+1}$ representation which we need to match with the flux truncation Hamiltonian in Eq.~\eqref{eq:fluxt}. 
For example, for the $M=4$ case, the decomposition of the full $\bigotimes_{m=1}^4 \mathbf{2}$ representation contains the irreducible $\mathbf{5}$ representation, on which the target Hamiltonian acts, as
\begin{align}\label{eq:4qubit-irrdecomp}
    \bigotimes_{m=1}^4 \mathbf{2} = \mathbf{1} \oplus \mathbf{1} \oplus \mathbf{3} \oplus \mathbf{3} \oplus \mathbf{3} \oplus \mathbf{5} 
\end{align}

\begin{figure}[ht]
    \centering
    $
    \ytableausetup{boxsize=0.75em}\ytableausetup{aligntableaux=top} \ydiagram[]{1}\otimes \ydiagram[]{1}\otimes \ydiagram[]{1}\otimes \ydiagram[]{1}
    =
    \ydiagram[]{2, 2}\oplus \ydiagram[]{2, 2} \oplus \ydiagram[]{3, 1} \oplus \ydiagram[]{3, 1} \oplus \ydiagram[]{4}
    $
    \caption{The Young tableau representation of the $SU(2)$ irreducible decomposition Eq.~\eqref{eq:4qubit-irrdecomp}. 
    We can embed the $U(1)$ Hamiltonian (Eq.~\eqref{eq:H-KS}) with the flux cutoff of $L=2$ into the symmetric representation $\mathbf{5}$ of $SU(2)$, which is represented as the last term of the right-hand side (the four boxes aligned in a single row). }
    \label{fig:ytab-su2}
\end{figure}

The simplifications where we start with $2M$ fermion qubits and end up with only $M$ qubits encodes an $M+1$ dimensional Hilbert space inside a $2^M$ dimensional Hilbert 
space. 
We have cut the number of qubits by half with the half-filling condition. 
Still, the dimension of the Hilbert space where our physics is encoded grows exponentially with the number of states. We will name this property an {\em exponential format}. 

The Hilbert space on which the $\mathbf{M+1}$ representation acts is spanned by the states that are totally symmetric on the flavor symmetry, i.e. the states 
\begin{align}\label{eq:sym-states}
    \ket{m} &\propto |\underbrace{\downarrow ... \downarrow}_m\underbrace{\uparrow ... \uparrow}_{M-m}\> + \text{permutations}
\end{align}
for $m = 0, ..., M$~\footnote{In the usual spin notation, we have that $m=j_z+j$, where $j$ is the spin of the representation and $j_z$ is the $z$ eigenvalue of angular momentum.}. 
The $+$permutations terms contain all the possible permutations of $m$ down spins and $M-m$ up spins. 
In this sense, if we gauge the permutation symmetry of the qubits, we get the unique representation of dimension $2L+1$. 
This can be justified by the gauging of the $U(M)$
flavor symmetry of the original D-theory formulation. 
The combinations in Eq.~\eqref{eq:gaugeinv} are 
invariant only under a $U(1)^M$ subgroup of $U(M)$ rather than the full $U(M)$. 
The individual $\sigma_z$ are linear combinations of the Cartan generators.
The commutant of $U(1)^M$ inside $U(M)$ also includes the permutations of the $U(1)$, which should also be gauged. 
It justifies this prescription for keeping only the symmetric states.

The ingredients for the construction of the qubit Hamiltonian are the Cartan-Weyl basis operators $\hat L_+ = \hat L_x + i \hat L_y$, $\hat L_-= \hat L_x - i \hat L_y$, and $\hat L_z$ in the $\mathbf{M+1}$ irreducible representation of $SU(2)$, i.e. the spin-$M/2$ operators.
This truncation of the $U(1)$ fields with the spin-$M/2$ operators is investigated by Zhang et al.~\cite{Zhang:2021dnz} to study the effect of this spin truncation to the BKT phase transition of the $O(2)$ model.
We can express these Cartan-Weyl basis operators in the full $\bigotimes_{m=1}^M \mathbf{2}$ representation, i.e. in the $M$-qubit representation as 
\begin{align}
   \hat U^\dagger &\rightarrow  \hat L_+ = \frac{1}{\sqrt{(M/2)(M/2 + 1)}}\sum_{m = 1}^M
  \sigma^+_m, \\ 
  \hat U &\rightarrow \hat L_- =
  \frac{1}{\sqrt{(M/2)(M/2 + 1)}}\sum_{m=1}^M  
  \sigma^-_m, \\
  \hat E & \rightarrow L_z = \frac{1}{2}\sum_{m = 1}^M \sigma^3_m
\end{align}
The normalization factors of $\hat L_\pm$ are given so that their actions on the $\ket{0}$ state match with those of the original $U$ operators, and hence the low spectra of the qubit Hamiltonian replicate those of the continuum Hamiltonian in the small $h$ region. 
The actions of the $\hat L_+$ and $\hat L_-$ operators on the symmetric states Eq.~\eqref{eq:sym-states} are to raise and lower them, respectively, as 
\begin{align}\label{eq:su2-ladders}
    &\hat L_+ \ket{m} = \frac{\sqrt{(m+1)(M - m)}}{\sqrt{(M/2)(M/2 + 1)}}\ket{m+1}, \nn 
    &\hat L_- \ket{m} = \frac{\sqrt{m(M - m + 1)}}{\sqrt{(M/2)(M/2 + 1)}}\ket{m-1}
\end{align}
and the $\hat L_z$ operator acts as the shifted number operator 
\begin{align}
    \hat L_z \ket{m} = \left(m - \frac{M}{2}\right)\ket{m}
\end{align}
Since the commutation relations of $\hat L_+$ and $\hat L_-$ with $\hat L_z$ are 
\begin{align}
    \left[\hat L_z, \hat L_+\right] = \hat L_+, \quad 
    \left[\hat L_z, \hat L_-\right] = -\hat L_-
\end{align}
which match with Eq.~\eqref{eq:E-U-algebra}, whereas the commutator of $\hat L_+$ and $\hat L_-$ is $\hat L_z$, which violates the unitarity $[U, U^\dagger] = 0$.
If we consider the mapping $E\rightarrow \hat L_z$, $U\rightarrow \hat L_+$, and $U^\dagger \rightarrow \hat L_-$, the Hamiltonian
\begin{align}\label{eq:H-qubit}
    \hat H_{\mathrm{qubit}} = \frac{g^2}{2} \hat L_z^2 - \frac{1}{2g^2}\left(\hat L_+ + \hat L_-\right)
\end{align}
has the same symmetry as the continuum Hamiltonian Eq.~\eqref{eq:H-KS} does (more precisely, the different pieces in the Hamiltonian have the same algebra). 

\subsection{Phase space considerations.}

One can try to understand this a little bit better from the point of view of Hamiltonian classical mechanics on the phase space of the cylinder and the sphere. The reason to do so is to understand better the relation between both dynamical systems. 

Basically, after turning to the problem of writing in terms of qubits and focusing  on the correct gauge invariant states, the original problem is reduced to the study of a single copy of the $SU(2)$ Lie algebra hiding in the big Hilbert space. It is the physics of this sub-Hilbert space that we want to analyze by  classical methods to get an intuition.

On the cylinder (the tangent bundle on the circle), we have variables $\alpha$ (the periodic variable) and $p_\alpha$, with the Poisson bracket
$\{\alpha,p_\alpha\}=1$. 
The cylinder Hamiltonian is
\begin{equation}
H= \frac 12 p_{\alpha}^2- h \cos(\alpha)
\end{equation}
By contrast, on the sphere we have a pair of spherical coordinates $\tilde\theta$, $\tilde\phi$, with $\tilde \phi$ periodic and 
with the Poisson bracket
$\{ \tilde \phi, \tilde\theta\}= A/\sin \tilde\theta$ (this is the inverse of the volume form in spherical coordinates, up to a rescaling factor, which we call $A$). 
The conjugate variable to $\tilde\phi$ is actually 
$p_{\tilde \phi} = \cos \tilde\theta/A$ rather than $\tilde\theta$. 
In the Cartesian coordinates, this is the
$z$ coordinate, and that is identified with $L_z$ after rescaling. 
On the other hand, $L^+\propto e^{i\tilde \phi} \sin\tilde \theta$, which is identified with $e^{i\tilde \phi} \sqrt{1-z^2} \sim e^{i\tilde \phi} \sqrt{1-L_z^2/L^2}$.
That is, when we take the classical identification $\alpha \equiv \tilde \phi $, which results from taking the classical periodicity of the variables into account, and include the constraint $L_x^2+L_y^2= L^2 -L_z^2$, where $L^2$ is a c-number, we find that the Hamiltonian actually takes the form
\begin{equation}
H= \frac 12 p_{\alpha}^2- h \sqrt{1-\frac{p_\alpha^2}{p_{\max}^2}}\cos\alpha
\end{equation}
For this to work, we need to have $p_\alpha= L_z$, 
$p_{\max}=L$ and $A= 1/p_{\max}=1/L$, so that $p_\alpha = p_{\max} \cos\tilde \theta$. 
The normalization of the naive kinetic term has been scaled to match what we need. 

We can now expand it in powers of $1/p_{\max}$ as 
\begin{equation}
H= \frac 12 p_{\alpha}^2- h \cos(\alpha)
+\frac h2 \frac{p_{\alpha}^2 }{p_{\max}^2} \cos(\alpha)+\dots\label{eq:higher_der}
\end{equation}
so when we take $p_{\max}\to \infty$, we recover the cylinder Hamiltonian. At finite $p_{\max}$, there are what should be interpreted as higher derivative corrections in the Hamiltonian. These are suppressed by the cutoff $p_{\max}$.
The quantization of this system leads to the quantum Hamiltonian Eq.~\eqref{eq:H-qubit}, provided that $1/g^4\propto h/{p_{\max}}$. 
Here, we need to remember that $L_+, L_-, L_z$ have roughly the same normalization. 
One can say that the quantity $1/p_{\max}$ is playing the role of $\hbar$ in a quantum expansion. This is also related to the volume of phase space, which is computed to be 
proportional to $p_{\max}\simeq (2L+1)$ in the Planck units. 

In a field theory setup, these higher derivative corrections are expected to be irrelevant perturbations, at least by naive power counting: they affect the UV dynamics but should flow to the same universality class in the infrared. 
These scale like the electric field squared, times the magnetic field squared (the naive plaquette).
In that vein, the low energy spectrum of Eq.~\eqref{eq:H-qubit} should converge to the low energy spectrum of the Kogut-Susskind Hamiltonian with the normalization of the equation Eq.~\eqref{eq:KShtetarep} when we take the cutoff to infinity as well. 

One can try to understand a similar idea for more general groups that are not just $U(1)$. 
As argued earlier, we should study the quantization of the Grassmannian ${\mathbb G(N,2N)}$, when we are discussing $U(N)$ links which are where the coherent states of \cite{Schlittgen:2000xg} take values. 
One would then try to understand how to take the semiclassical double scaling limit correctly to get a {\em cylinder over $U(N)$}, namely, the tangent space of $U(N)$ as a phase space, with a Hamiltonian and a parameter playing the role of $p_{\max}$. 
This type of analysis is beyond the scope of the present work.

\subsection{Exponential Formats\label{sec:exp-formats}}

The Hamiltonians Eq.~\eqref{eq:H-KS} and Eq.~\eqref{eq:H-qubit} are not exactly the same due to the difference between the coefficients of the actions of the raising and lowering operations. 
In the flux truncation, they are always constant (normalized to $1$), whereas those in $SU(2)$ are not constant but rather depend on the target state as we saw in Eq.~\eqref{eq:su2-ladders}. 
This is also true in the classical limit described by Eq.~\eqref{eq:higher_der}, where there are higher derivative corrections to the Hamiltonian. 
Also, remember that the Hamiltonian Eq.~\eqref{eq:H-qubit} acts on a Hilbert space of large dimension $2^M$, but that only the states in the $\mathbf{M+1}$ irreducible representation matter and are invariant under all constraints.
We dubbed this property as being an exponential format, where the dimension of the Hilbert space grows exponentially in the number of states we need.

We can try to do better at the level of matching the operators in the subspace of interest in the total Hilbert space, by adding corrections to the operators getting rid of the differences below the finite cutoff. 
This should be equivalent to adding (or depending on the point of view, subtracting) irrelevant operators to compensate for the differences in the formulation. 

To construct the $U$ and $U^\dagger$ with our qubit construction, we may use one of the two ansatze:
\begin{align}
    \hat U' = \sum_{k=0}^{M/2-1} a_k \hat L_z^k \hat L_+ \hat L_z^k, 
    \quad 
    \hat U'' = \sum_{k=0}^{M/2-1} b_k (\hat L_+ \hat L_-)^k \hat L_+
\end{align}
The first one, for what we call $\hat U'$ can also be thought of as having $L_z/L_{\max}$ corrections to $U^+$, as one would expect from the higher derivative expansion Eq.~\eqref{eq:higher_der}. As such, the coefficients should be  suppressed by the powers of $1/L_{\max}^{2k}$, up to normal ordering ambiguities.

One can compute the coefficients $a_k$ or $b_k$ so that the action of $\hat{U}'$ or $\hat U''$ is the same as $U$ in Eq.~\eqref{eq:H-KS}. 
For the first $\hat U'$ for example, the action of each $\hat L_z^k \hat L_+ \hat L_z^k$ operator is as 
\begin{align}
    &\hat L_z^k \hat L_+ \hat L_z^k\ket{m}
     = 
    A_{mk} \ket{m+1}
\end{align}
where $A_{mk}$ ($m, k = 0, 1, ..., M/2-1$) is a $(M/2) \times (M/2) $-dimensional matrix with the elements of 
\begin{align}
    A_{mk} &= \left[\left(m -\frac{M}{2}+ 1\right)  \left(m - \frac{M}{2}\right)\right]^k \nn 
    &\quad \times \frac{\sqrt{(m+1)(M - m)}}{\sqrt{(M/2)(M/2 + 1)}}
\end{align}
The coefficients $a_k$ can be computed by solving the linear equation 
\begin{align}
    \sum_{k=0}^{M/2-1} A_{mk}a_k = 1
\end{align}
Similar procedure can be taken to find the coefficients for the other case of $\hat U''$ as well. 
Note that the operators $\hat{U}'$ and $\hat{U}''$ constructed with the appropriate coefficients $a_k$ and $b_k$ are identical in the space spanned by the states Eq.~\eqref{eq:sym-states}, even though they are not in the full $M$-qubit space.

Since the values of $A_{mk}$ and $B_{mk}$ grow exponentially with $k$, the values of $a_k$ and $b_k$ are expected to decay exponentially for higher $k$ terms.  
Indeed, \cite{Zhang:2021dnz} numerically demonstrates that $a_k$ shows the behavior of exponential decay with $k$. 
For small cutoff $L=1, 2, 3$, the $\hat U'$ operator can be constructed as 
\begin{align}
    \hat U'_{L=1} &= \hat L_+\nn
    \hat U'_{L=2} &= \hat L_+ + \left(-\frac{1}{2} + \frac{\sqrt{6}}{4}\right) \hat L_z \hat L_+ \hat L_z  \nn
    \hat U'_{L=3} &= \hat L_+ + \left(-\frac{2}{3}- \frac{\sqrt{2}}{12}+ \frac{3\sqrt{30}}{20}\right) \hat L_z \hat L_+ \hat L_z \nn
    &\quad + \left(\frac{1}{12} +  \frac{\sqrt{2}}{24} - \frac{\sqrt{30}}{40}\right) \hat L_z^2 \hat L_+ \hat L_z^2 
\label{eq:Form1}
\end{align}
and the $\hat U''$ operator can be constructed as 
\begin{align}
    \hat U''_{L=1} &= \hat L_+\nn
    \hat U''_{L=2} &= \left(3 \sqrt{\frac{3}{2}}-2\right) \hat L_+ + \left(\frac{1}{2} - \frac{\sqrt{6}}{4}\right) \hat L_+ \hat L_- \hat L_+\nonumber \\
    \hat U''_{L=3} &= \left(5-9 \sqrt{\frac{6}{5}}+5 \sqrt{2}\right) \hat L_+ \nn 
    &\quad + \left(-\frac{3}{4} - \frac{11\sqrt{2}}{12} + \frac{9\sqrt{30}}{20}\right) \hat L_+ \hat L_- \hat L_+\nn
    & \quad + \left(\frac{1}{12} +  \frac{\sqrt{2}}{24} - \frac{\sqrt{30}}{40}\right) \hat (L_+ \hat L_-)^2 \hat L_+
\label{eq:Form2}
\end{align}

For the expressions of $\hat U'$, notice that numerically we have 
$-1/2+\sqrt 6/4\sim 0.11$, and $-2/3-\sqrt{2}/12+3\sqrt{30}/20\sim 0.049\sim (2/3)^2 *(0.11)$ is
roughly the suppression one would expect in terms of the $1/p_{\max}^2$
counting.

The point is that if our goal is to produce the flux truncated
Kogut-Susskind Hamiltonian on the nose, it can be done. 
Ideally, one would actually use $L^+, L^-$ instead and try to argue that one is in the same universality class.
The main reason is that the Hamiltonian Eq.~\eqref{eq:H-qubit} is made of sums of products of only two sigma matrices. 
These can be readily implemented as 2 qubit gates, perhaps
with some swaps of qubits. 
Therefore, it provides a more efficient
implementation on a NISQ device, where reducing the number of total
gate operations per qubit is essential to get to a result that one can trust, before one loses coherence on the device.

\subsection{Linear Formats and Sparsity}
The formulation we have used to construct the qubit Hamiltonian with
links in Eqs.~\eqref{eq:Form1} and \eqref{eq:Form2} requires at least
$M = 2 L$ qubits,  discarding all
other representations in the irreducible decomposition of
$\bigotimes_{\alpha=1}^M \mathbf{2}$ than the $\mathbf{M+1}$
representation.  
For large $L$, the Hilbert space grows exponentially in $L$ losing the {\em quantum advantage} locally~\footnote{The exponential formats still enjoy the quantum advantage once we turn to the theory with $d > 0$ spatial dimensions}. 
We now introduce another qubit
representation with which one can store information with only a logarithmic number of qubits, using
$M$ qubits to represent the $M+1$-dimensional Hilbert space. 

One needs only $n_{\min} = \lceil \log_2 (M+1) \rceil$ qubits by keeping the other representations by mapping the
$\ket{0}, \ket{1}, ..., \ket{M}$ states to the computational basis (i.e. the eigenbasis of $\sigma^3$) states as
\begin{align}
    \ket{m} \mapsto \ket{b_{n_{\min}}} \otimes \ket{b_{n_{\min}-1}} \otimes \cdots \otimes \ket{b_1} \otimes \ket{b_0}
\end{align}
where $b_{n_{\min}} b_{n_{\min}-1} b_1 b_0$ is the binary
representation of the integer $m$. 
In this encoding, the dimension of the Hilbert
space in which we embed our problem grows linearly with the dimension
of the Hilbert space we want to encode.
We would call this a \textit{linear format}.
Notice that this setup starts with the truncation and tries to
fit it into a Hilbert space in an arithmetic way without starting with
the symmetry algebra first. It is more economical in terms of qubits,
but the generalization  to non-Abelian fields for even a polynomial format is not straightforward,
and even if possible presents a challenging research project in qubit algebra~\cite{Takahashi_2009}.

We begin by introducing the $M$-bit quantum adder~\cite{Takahashi_2008}:
\begin{align}
    A &= \sigma^+_0 + \sigma^+_1\sigma^-_0 + \sigma^+_2 \sigma^-_1 \sigma^-_0 +\nn 
    &\quad \cdots + \sigma^+_{n_{\min}-1} \sigma^-_{n_{\min}-1} \cdots \sigma^-_1 \sigma^-_0
\end{align}
which maps the computational basis states as $\ket{m} \rightarrow \ket{m+1, n_{\min}}$ mod $2^M$.
Then, the adder can be modified to represent the raising operator $\hat U_{\min}$ replacing the mod by annihilation for the highest state as  $\hat U_{\min} \ket{M} = 0$.  
To do this in general, we multiply the projector $P$ from right to $A$, where $P$ is defined to act as the identity for the $\ket{0}, ..., \ket{M-1}$ states and vanish at least the state $\ket{M}$ and possibly also the higher states. 
For $M = 2 (L = 1)$, for example, 
using the $3$-dimensional subspace of the $2$-qubit space with a mapping given as 
\begin{align}
    \ket{0}\mapsto \ket{00}, \ket{1}\mapsto \ket{01}, 
    \ket{2}\mapsto \ket{10}
\end{align}
The $\hat U_{\min}$, $\hat U_{\min}^\dagger$, and the corresponding $\hat E_{\min}$ operator are expressed as 
\begin{align}
    \hat U_{\min} &= A P  = \frac{1}{2}\sigma^+_0
    +\sigma^+_1\sigma^-_0
    +\frac{1}{2}\sigma^3_1\sigma^+_0
    \\
    \hat U_{\min}^\dagger &= P^\dagger A^\dagger = \frac{1}{2}\sigma^-_0
    +\sigma^-_1\sigma^+_0
    +\frac{1}{2}\sigma^3_1\sigma^-_0
    \\
    \hat E_{\min} &=
    -\frac{1}{2} \sigma^3_1
    -\frac{1}{2}\sigma^3_1\sigma^3_0
\end{align}
given that one of the possible choices of the projector $P$ is 
\begin{align}
    P = I - \frac{(I - \sigma^3_1)(I - \sigma^3_0)}{4}
\end{align}
For another example of $M = 4 (L = 2)$, 
using the $5$-dimensional subspace of the $3$-qubit space with a mapping given as 
\begin{align}
    &\ket{0}\mapsto \ket{000}, \; \ket{1}\mapsto \ket{001}, \; 
    \ket{2}\mapsto \ket{010}, \nn 
    &\ket{3}\mapsto \ket{011},  \; 
    \ket{4}\mapsto \ket{100}
\end{align}
the $\hat U_{\min}$, $\hat U_{\min}^\dagger$, and the corresponding $\hat E_{\min}$ operator are expressed as 
\begin{align}
    \hat U_{\min} &= A P \nn
    &= \frac{1}{2}\sigma^+_0
    +\frac{1}{2}\sigma^+_1\sigma^-_0
    +\frac{1}{2}\sigma^3_2\sigma^+_0\nn 
    &\quad +\frac{1}{2}\sigma^3_2\sigma^+_1\sigma^-_0
    +\sigma^+_2\sigma^-_1\sigma^-_0\\
    \hat U_{\min}^\dagger &= P^\dagger A^\dagger 
    \nn 
    &= \frac{1}{2}\sigma^-_0
    +\frac{1}{2}\sigma^-_1\sigma^+_0
    +\frac{1}{2}\sigma^3_2\sigma^-_0\nn 
    &\quad 
    +\frac{1}{2}\sigma^3_2\sigma^-_1\sigma^+_0
    +\sigma^-_2\sigma^+_1\sigma^+_0\\
    \hat E_{\min} &=
    -\frac{1}{4} \sigma^3_1
    + \frac{1}{4}\sigma^3_1\sigma^3_0
    -\frac{1}{2}\sigma^3_2 
    -\frac{1}{2}\sigma^3_2\sigma^3_0\nn
    &\quad -\frac{3}{4}\sigma^3_2\sigma^3_1
    -\frac{1}{4}\sigma^3_2\sigma^3_1\sigma^3_0
\end{align}
given that one of the possible choices of $P$ is 
\begin{align}
    P = \frac{I - \sigma^3_2}{2}
\end{align}
These operators also satisfy the commutation relations Eq.~\eqref{eq:E-U-algebra}, so the Hamiltonian constructed from these operators
\begin{align}
   \hat H_{\min} = \frac{g^2}{2} \hat E_{\min}^2 - \frac{1}{2g^2} \left(\hat U_{\min} + \hat U_{\min}^\dagger\right)
\end{align}
preserves the original symplectic algebra.  
Drawing on the extensive literature on efficient and robust
quantum arithmetic~\cite{Takahashi_2009} should help in designing optimal circuits for this linear formation. 


\section{Spectral Matching of D-theory
  Truncation\label{sec:spectral-matching}}
In this section, we discuss and numerically compare the low spectra of the $0+1$-D quantum rotor Hamiltonian with the $U(1)$ symmetry defined in Eq.~\eqref{eq:H-KS} with those with a small flux cutoff $L$, a discretization of the group manifold of $U(1)$ to $\mathbb Z_N$ (clock model), and the spin operators $L_\pm$ as $\hat U$ operators (quantum link model) constructed with $M$ flavor qubits. \\

\begin{figure}[t]
    \centering
    \subfloat[]{
   \includegraphics[width=.46\linewidth]{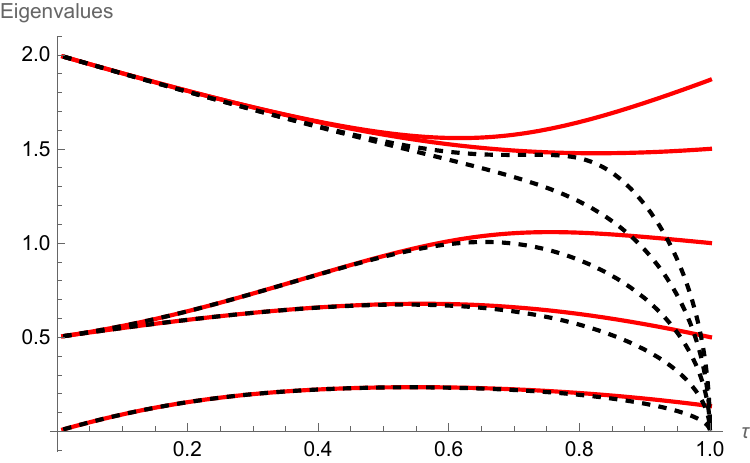}
   \quad 
   \includegraphics[width=.46\linewidth]{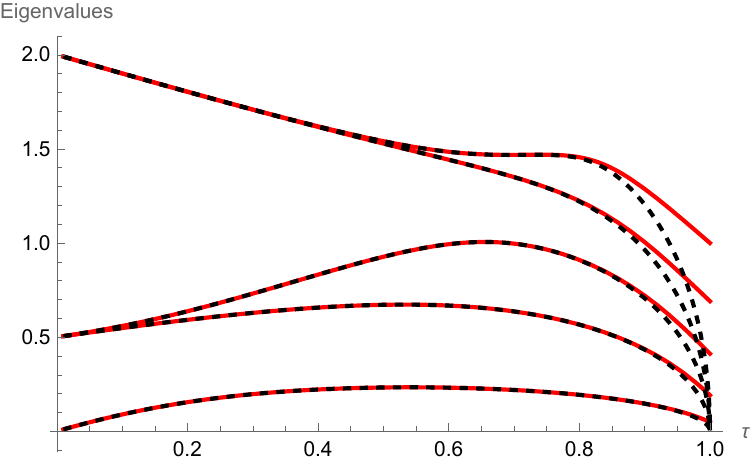}
   }\\
   \subfloat[]{
   \includegraphics[width=.46\linewidth]{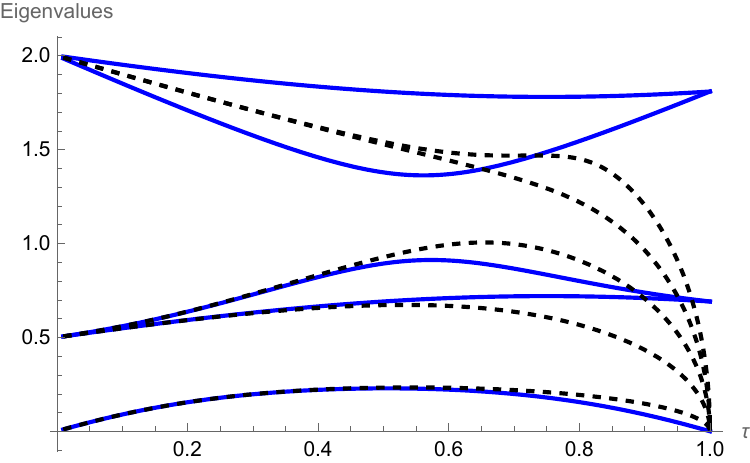}
   \quad 
   \includegraphics[width=.46\linewidth]{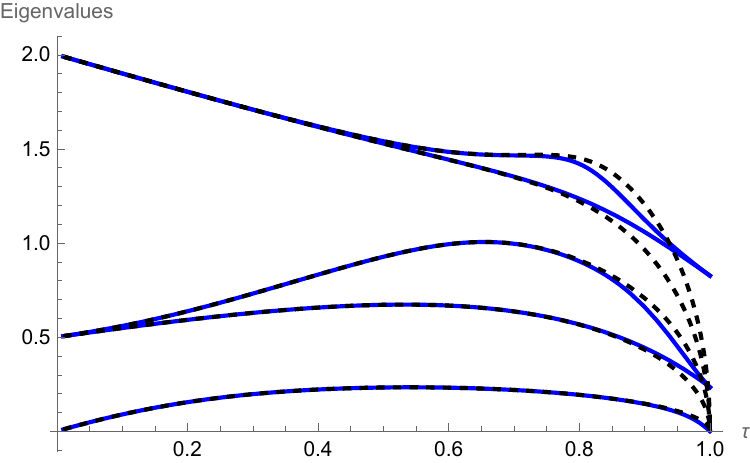}
   } \\ 
   \subfloat[]{
   \includegraphics[width=.46\linewidth]{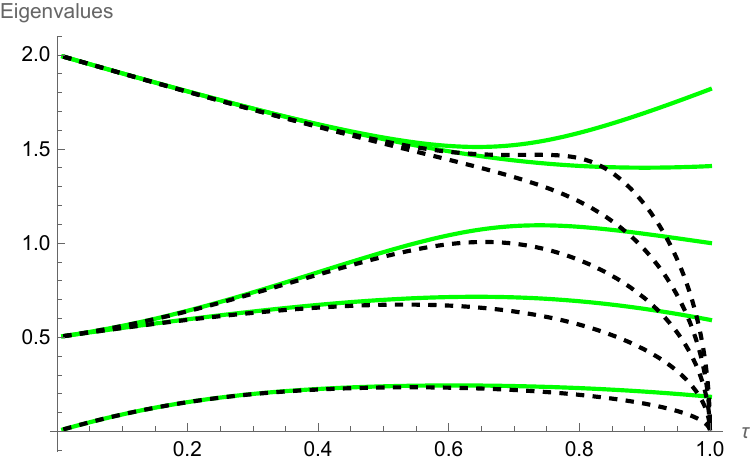}
   \quad 
   \includegraphics[width=.46\linewidth]{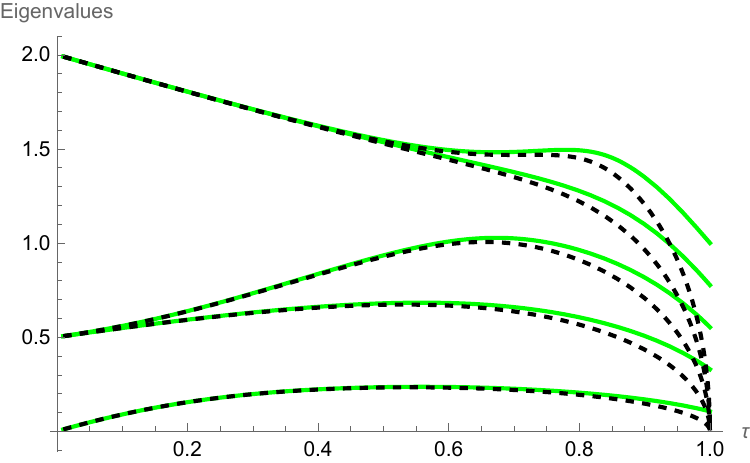}
   }
    \caption{The lowest five eigenvalues as the functions of $\tau$ for the quantum rotor Hamiltonian computed with (a) a small flux cutoff with (left)~$L=2$ and (right)~$L=4$, (b) the clock model discretization with (left)~$\mathbb Z_5$ (right)~$\mathbb Z_9$, and (c) the spin operators $\hat L_\pm$ as the $\hat U$ operators with (left)~$M = 2L = 4$ and with (right)~$M = 2L = 8$, compared with the spectrum of the exact Hamiltonian (black dashed curves). 
    }
    \label{fig:flux-clock-spectra}
\end{figure}

First, we compare the spectra with the very small cutoff giving the five-dimensional Hilbert space and those with a slightly larger cutoff with the nine-dimensional Hilbert space (Fig.~\ref{fig:flux-clock-spectra}) with the exact spectra. 
We define a new coefficient $\tau$ parameterizing the inverse coupling $h$ as $h = \tau/(1-\tau)$ allowing us to plot the whole $h \in [0, \infty)$ with the finite $\tau \in [0, 1)$, besides rescaling the Hamiltonian by $\times (1-\tau)$. 
We can see from the figures that for the strong coupling region (small $h$), we do not need a large cutoff $L$ to reach the precise solution for all of the truncation approaches, whereas we do need large $L$ for the weak coupling. 
We can also observe that the lower eigenenergies converge to the exact values faster than the higher energies.
Let us also note that the spectrum of the clock model deviates from the exact spectrum with smaller $h$ than that with the flux cutoff does. 

\begin{figure}[ht]
    \centering
    \subfloat[]{
   \includegraphics[width=.8\linewidth]{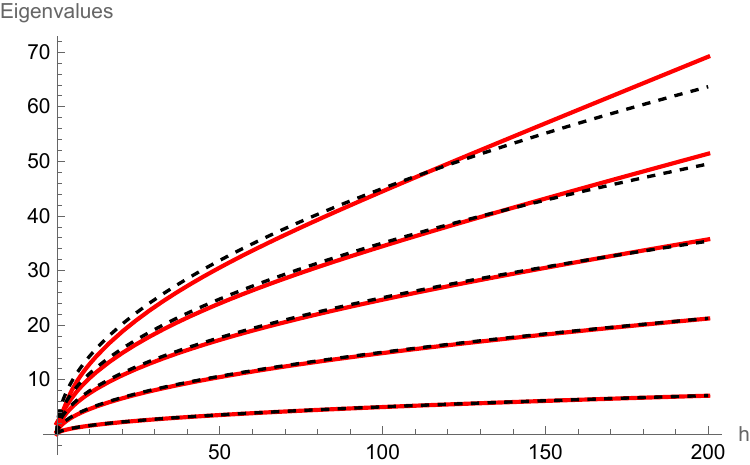}
   }\\
   \subfloat[]{
   \includegraphics[width=.8\linewidth]{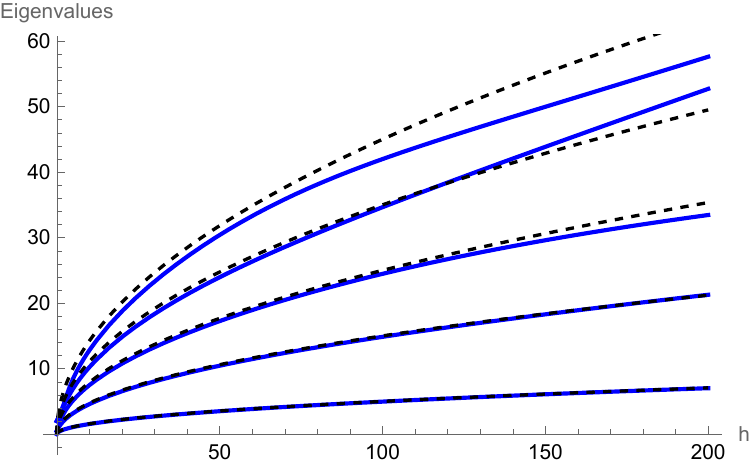}
   }\\
   \subfloat[]{
   \includegraphics[width=.8\linewidth]{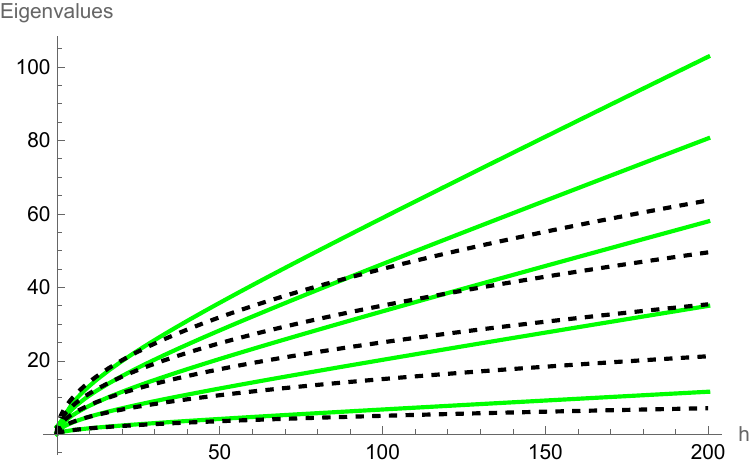}
   }
    \caption{The lowest five eigenvalues of the quantum rotor Hamiltonian as functions of $h$ computed with (a)~a flux cutoff with $L=10$, (b)~the group manifold discretization to $\mathbb Z_{21}$ group, and (c)~the spin operators with $M=20$, compared with the spectrum of the QHO (black dashed curves). 
    }
    \label{fig:flux-clock-qho}
\end{figure}

It is also worth noting that the quantum rotor can be locally approximated as the quantum harmonic oscillator (QHO) around $\theta = 0$. 
By expanding the $\cos\theta$ term by $\theta$, we can decompose the Hamiltonian to the non-perturbed QHO part $H_0 = p^2/2 + h \theta^2/2$ with the momentum $p = -i\dd_\theta$ and the perturbation term $H_1 = h(- \theta^4/4! + \theta^6/6! - \theta^8/8! + \cdots)$ which is the higher-order terms of the cosine. 
In the large $h$ region, the low spectra tend to condensate around $\theta = 0$, which makes the perturbation $H_1$ and the periodic boundary conditions of $\theta$ trivial, so they should exhibit the even-spaced spectra as well as the well-known QHO spectra with $E_n = \sqrt{h}(n + 1/2)$ for $n = 0, 1, 2, ...$. 
To see if the two truncations can reproduce this QHO-like behavior with large $h$, we compare the spectra of the Hamiltonians with the truncations with the spectra of the QHO (Fig.~\ref{fig:flux-clock-qho}). 
These QHO solutions correspond to the topologically trivial trajectories with zero winding, whereas the topologically non-trivial trajectories start to appear as nonperturbative effects in the small $h$ region once we take the path integral for quantization~\cite{Smilga:2001ck}.
Since we need to simulate the behavior with large $h$, we use larger truncations giving the Hilbert space with dimension 21. 
As seen in the figure, the flux truncated Hamiltonian and the clock model well reproduce the QHO spectra in the large $h$ regions until they start to experience non-negligible errors due to the truncations. 
However, the spectra computed with the spin truncation are completely off from the QHO spectra, which can be expected given that the dominating potential term $2 - \hat U - \hat U^\dagger = 2 - \hat L_x$ in the large $h$ region has the even spaced eigenvalues of $2 - 2m/\sqrt{(M/2)(M/2+1)}$ with $m = -M/2, -M/2+1, ..., M/2-1, M/2$ which grow linearly with $h$, instead of its square root. 
This indicates we need the corrections as we proposed in Sec.~\ref{sec:exp-formats} for $h$ very large.
At intermediate $h$, we cannot neglect the term with $L_z^2$ and the agreement should be better.

\begin{figure}[t]
    \centering
    \subfloat[]{
    \includegraphics[width=.9\linewidth]{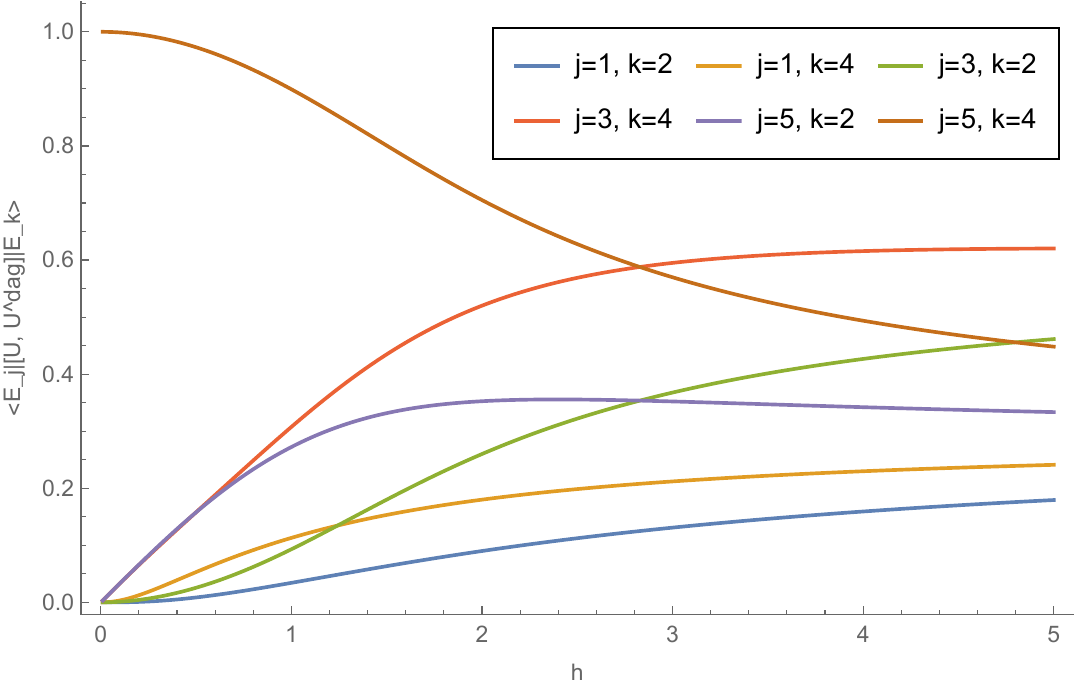}
    }\\
    \subfloat[]{
    \includegraphics[width=.9\linewidth]{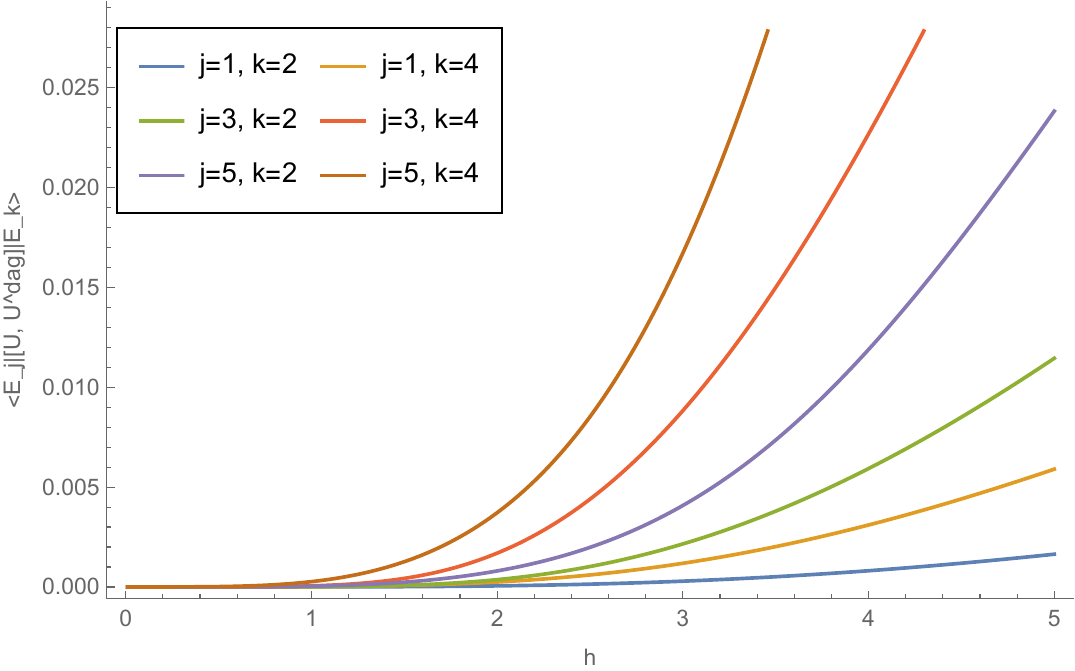}
    }
    \caption{The non-zero matrix elements in the eigenbasis as the functions of $h$ for low spectrum of Eq.~\eqref{eq:commutator-matrix} with the flux cutoff (a)~$L=2$ and (b)~$L=4$. }
    \label{fig:commutator-mat-elem}
\end{figure}

To evaluate the performance of the flux truncation, we can also look at the breaking of the unitarity constraint in the eigenbasis for the low spectrum
\be\label{eq:commutator-matrix}
\< E_{j} | [U,U^\dag ] |E_k\>
\ee
which is, without the truncation or with the clock model discretization, exactly zero. 
For $j$ and $k$ with the same parity, the matrix elements are all zero. 
The non-zero elements (i.e. $j$ and $k$ with the different parities) for small $j$ and $k$ of this matrix as functions of $h$ with a small cutoff ($L=2$ and $L=4$) are as Fig.~\ref{fig:commutator-mat-elem}, demonstrating that the effects of the breaking on the low energy states are small for smaller $j$ and $k$, for smaller coupling $h$, and larger cutoff $L$. 
This behavior of the breaking of the zero commutator validates that the flux-truncated Hamiltonian describes the effective theory of the exact $U(1)$ quantum rotor in the small $h$ region.

\section{Applications to 1 + 1 Field Theories. }
\label{sec:Applications}

We can think of simple $1+1$-D models to which our scheme can be applied for the simulations of their dynamics. 
An interesting choice is the Sine-Gordon Model with the Lagrangian
\begin{align}
    \mathcal L_{\text{SG}} = \frac{1}{2} \dd_\mu \phi \dd^\mu \phi  + \frac{m^2}{\beta^2}  (1 - \cos(\beta \phi))
\end{align}
This is an intriguing exactly integrable theory with a strong-weak
S-duality to the massive-Thirring model as shown in~\cite{Coleman1975} demonstrating the fermionic excitations in the massive-Thirring model correspond to the solitons in the Sine-Gordon model. 
We note that the simulation of the massive-Thirring model on a quantum circuit is studied in~\cite{Mishra:2019xbh}. 
Both forms could be formulated for qubits with complimentary regions to find a common parameter space exhibiting the duality.

Given that the latticized derivative $\dd_\mu \phi := \frac{1}{a} (\phi(x + \mu) - \phi(x))$ is small in the low-energy range where $a$ is the lattice spacing, we can map the conjugate momentum field as $\pi(x) = \dd_0 \phi (x) \rightarrow E_x$ and the compactified field $\exp(i\beta \phi(x)) \rightarrow U_x$ leading to the Hamiltonian as 
\begin{align}\label{eq:SG-H}
     H_{\text{SG}} &=\frac{1}{2} \sum_x [  E^2_x + h  (2 - U_x - U^\dag_x) ] \nn 
&\quad +
J \sum_{\<x,y\>} (2 - U_x  U^\dag_y - U_y U^\dag_x)
\end{align}
where $h = m^2/\beta^2$ and $J = 1/2a^2\beta^2$. 
We fix the lattice spacing to be $a = 1$ from now on. 
We note that for $h = 0$, this is the classical XY model or the integer-spin XX chain which has been numerically studied in more detail by Zhang, Meurice and Tsai~\cite{Zhang:2021dnz} with tensor
networks and the range of truncation of $L = 1,2,3,4$. 
Among the interesting observations, while all preserve the gapless phase, the model has an infinite-order Gaussian transition for $L=1$ and only for $L \ge 2$ has a BKT transition.
This is a nice example of how physics can depend crucially on the size of truncation.
It has been known that the Sine-Gordon model effectively describes the vortices in the XY model with $\beta$ corresponding to the inverse temperature, and its BKT transition is well-studied from its renormalization flow on the Sine-Gordon side~\cite{Gogolin:2004rp}.

\subsection{Real-time evolution}
It is plausible that for the small $M$ qubit formulation, we can
explore the small $h$ and $J$ region. 
This indicates that the qubit Hamiltonian is expected to be the effective theory of the exact Hamiltonian in the low-temperature limit. 
Here, we give the smallest truncation $L=1$, and simulate the
real-time evolution of a state under the Hamiltonian with quantum
circuits defined on six lattice sites.
For $L=1$, the Hamiltonian $ H_{\text{SG}}$ can be represented with sigma matrices on a $1+1$-D lattice (ignoring the constant) as 
\begin{align}
    H_{\text{SG}}
 &= \frac{1}{8}\sum_x \sigma^3_{x, 1} \sigma^3_{x, 2} - \frac{h}{2\sqrt{2}} \sum_{x} ( \sigma^1_{x, 1}  +\sigma^1_{x, 2}) \nn 
 &\quad - \frac{J}{2} \sum_{\<x, y\>} \sum_{i, j = 1}^2  \left(\sigma^+_{x, i} \sigma^-_{y, j}  +  \sigma^-_{x, j} \sigma^+_{y, i}\right)
\end{align}

\begin{figure}
    \includegraphics[width=\linewidth]{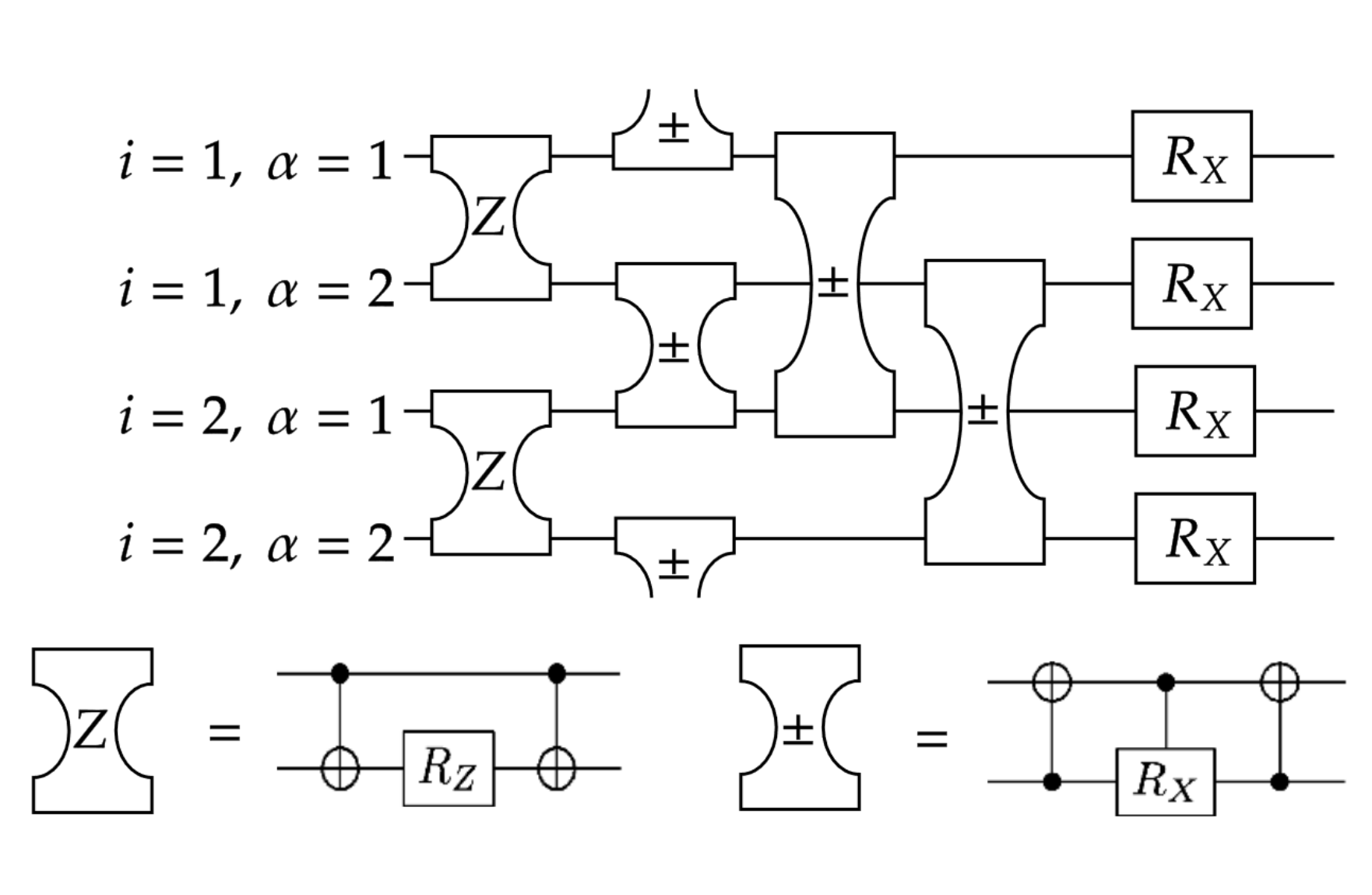}
    \caption{The quantum circuit of the single Trotter step for the Sine-Gordon model for two lattice sites with the periodic boundary conditions (top). 
    The index $i$ represents the position of the link and $\alpha$ represents the flavor. 
    The circuit components $Z$ and $\pm$ represent the operator $\exp( -i\Delta t\frac{1}{8} \sigma^3_{x, 1} \sigma^3_{x, 2})$ (can be realized as bottom left) and $\exp( -i \Delta t\frac{ J}{2} \left(\sigma^+_{x, i} \sigma^-_{y, j}  + \sigma^-_{x, i} \sigma^+_{y, j}\right))$ (bottom right), respectively. }
    \label{fig:circ-SG}
\end{figure}

\begin{figure*}[t]
    \centering
    \includegraphics[width=\linewidth]{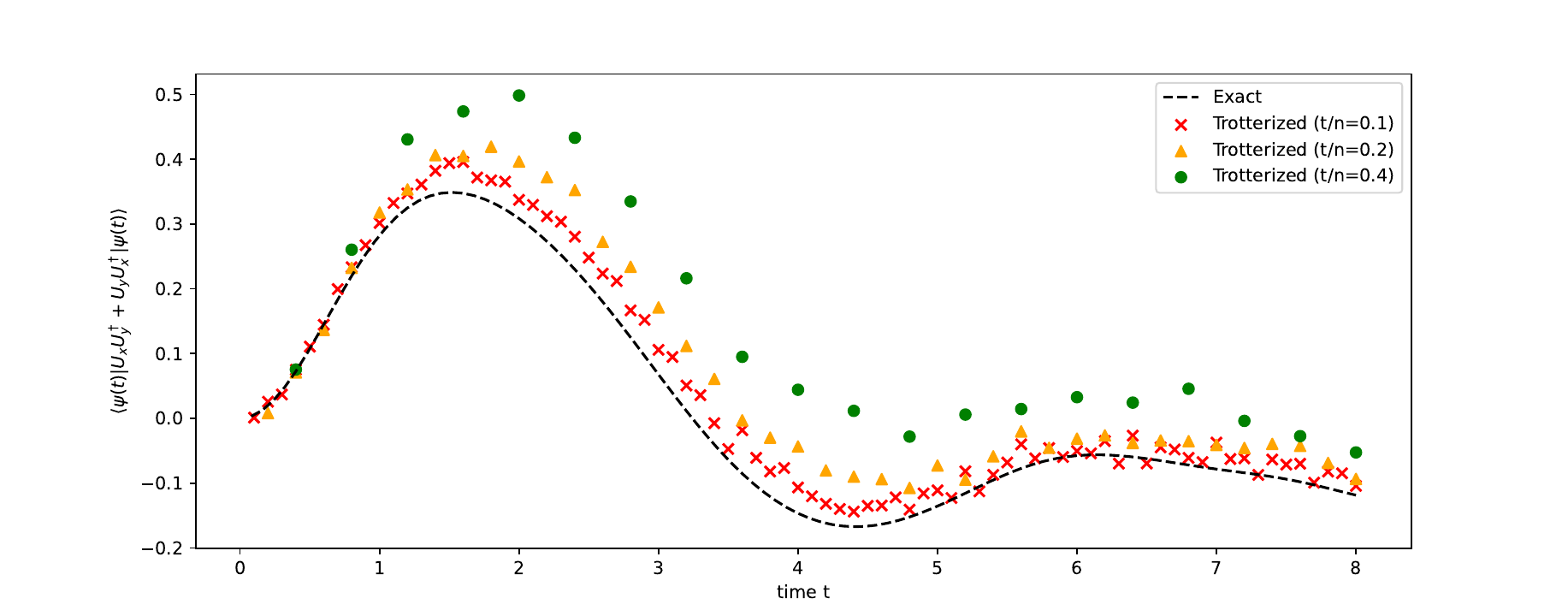}
    \caption{The simulated time evolution of the value $\frac{1}{2} \expval{U_x^\dagger U_y + U_x U_y^\dagger}$ with $\ket{\psi(t = 0)} = \ket{00...000}$ with different time intervals ($\Delta t = t/n = 0.1, 0.2, 0.4$). }
    \label{fig:trotter-sg}
\end{figure*}

Since the first electric term, the second potential term, and the third interaction term of $H_{\text{SG}}$ do not commute with each other, we use the Trotter-Suzuki approximation to simulate the time-evolution operator $\exp(-iH_{\text{SG}}t)$ on a quantum circuit with a small time step $\Delta t \equiv t/n$ with $n \gg 1$: 
\begin{align}\label{eq:trotter-SG}
    &\exp( -it H_{\text{SG}})\nn
    &\approx 
    \bigg[\prod_x \exp(- i \Delta t \frac{1}{8} \sigma^3_{x, 1} \sigma^3_{x, 2}) \nn 
    &\quad \times \prod_x \exp( i \Delta t \frac{h}{2\sqrt{2}} ( \sigma^1_{x, 1}  +\sigma^1_{x, 2}) ) \nn 
    &\quad \times \prod_{\<x, y\>}\prod_{i, j = 1}^2 \exp( i\Delta t \frac{J}{2}  \left(\sigma^+_{x, i} \sigma^-_{y, j}  +  \sigma^-_{x, j} \sigma^+_{y, i}\right))\bigg]^n
\end{align}
We call the product inside the bracket a Trotter step, and the realization of the single Trotter step is depicted in Fig.~\ref{fig:circ-SG}. 
Each unitary rotation component can be realized with simple one or two-qubit quantum operations. 
Eq.~\eqref{eq:trotter-SG} means that we can approximate the time evolution $e^{\large -it H_{\text{SG}}}$ on a quantum circuit by iterating the Trotter step circuit many times with a small time step of $\Delta t$. 
To test the reliability of the approximated time-evolution operator on a quantum circuit, we construct and simulate the circuit using \textit{qiskit} with six lattice sites, i.e. twelve qubits. 
We pick the two-point correlation function in the spatial dimension (specifically, the left-most lattice site $x$ and the middle point $y$) as the physical quantity to be measured from the circuit:
\begin{align}
    \expval{\cos(\phi(x) - \phi(y))} &\approx \frac{1}{2}\expval{U_x^\dagger U_y + U_x U_y^\dagger}\nn 
    &=
    \frac{1}{4}\sum_{\alpha, \beta = 1}\expval{\sigma^1_{x, \alpha}\sigma^1_{y, \beta} + \sigma^2_{x, \alpha}\sigma^2_{y, \beta}}
\end{align}
Since the observable $U_x^\dagger U_y + U_x U_y^\dagger$ can be expressed as a sum of the product of two sigma matrices with our construction, it can be measured by simple two-qubit measurements on the quantum circuit.  
We measure each two-qubit Paulis with 4096 runs of the circuit to approximate the expectation value. 
We choose the parameters of the Hamiltonian to be $h = 1$ and $J = 1$, and the state $\ket{\psi} = \ket{00...0}$ corresponding to the state whose sites all have the flux of $\ell=-1$ as the initial state which can be easily realized on a quantum circuit. 
The result of the simulation is Fig.~\ref{fig:trotter-sg}, from which we can see that if the value of the time interval $\Delta t$ is small enough ($\approx 0.1$), the quantum circuit well approximates the exact time evolution. 
\\

\subsection{Gapped/gapless phase transition}
As we mentioned above, the interesting physical feature to investigate for these $1+1$-D $U(1)$ models in the continuum is the BKT transition. 
Since the BKT transition is due to the topological defects in the model, it is regarded as a topological phase transition.  
For example, the transition in the $2$-D classical XY model can be explained as the confinement/deconfinement phase transition of the vortex-antivortex pairs. 
The topologically ordered phase is gapped, i.e. has a finite correlation length $\xi$. 
The topological phase transition closes this mass gap and allows the system to have massless Nambu-Goldstone excitations, and hence this other phase is critical and has an infinite correlation length. \\

We can observe this gapped/gapless transition by computing the entanglement entropy of the ground state.
Here we consider two entropy measures. 
The von Neumann entanglement entropy of the ground state for the subsystem $A$ is defined as $S_{A} := -\Tr[\rho_{A} \log \rho_{A}]$ where $\rho_{A}$ is the density matrix of the ground state in the subsystem $A$ defined as $\rho_{A} = \Tr_{A^c}[\rho]$ with the density matrix of the ground state $\rho$ in the total system $A \cup A^c$. 
The $\alpha$-Renyi entanglement entropy is defined as $S_A^{(\alpha)} = \frac{1}{1-\alpha} \log \Tr[\rho_A^\alpha]$, and it is related to the von Neumann entropy by $\lim_{\alpha\rightarrow 1}S_A^{(n)} = S_A$ (so-called the \textit{replica trick}). 
It is proven by Hasting~\cite{Hastings:2007} that the entanglement entropy of the ground state of $1+1$-dimensional gapped systems obeys the area law, i.e. it is bounded from above by a constant which is independent of the subsystem size $n$. 
On the other hand, the entanglement $S_{A}$ needs a logarithmic correction $\log(n)$ in a gapless phase or a critical point in the thermodynamic limit, specifically it is proven for $1+1$-dimensional systems by Calabrese and Cardy~\cite{Calabrese:2004eu} by means of the two-dimensional conformal symmetry. 
For the finite volume cases, they prove that the Renyi entropy for the $1+1$-D quantum system with conformal symmetry on the finite lattice is (ignoring the constant term)
\begin{align}\label{eq:ee-critical-finite-Renyi}
    S_A^{(\alpha)} = -\frac{\alpha+1}{\alpha}\frac{c}{6}\log(\frac{N}{\pi a}\sin(\frac{\pi \alpha}{N}))
\end{align}
which converges to the von Neumann entropy in $\alpha \rightarrow \infty$ as 
\begin{align}\label{eq:ee-critical-finite-vN}
    S_{A} = \frac{c}{3} \log(\frac{N}{\pi a}\sin(\frac{\pi n}{N}))
\end{align}
In the thermodynamic limit $N\rightarrow \infty$, it converges to the logarithmic correction of $\log(n/a)$.
$c$ is the central charge of the conformal theory in the same universality class as the quantum system. 
\\

\begin{figure}[t]
    \subfloat[]{
    \includegraphics[width=.9\linewidth]{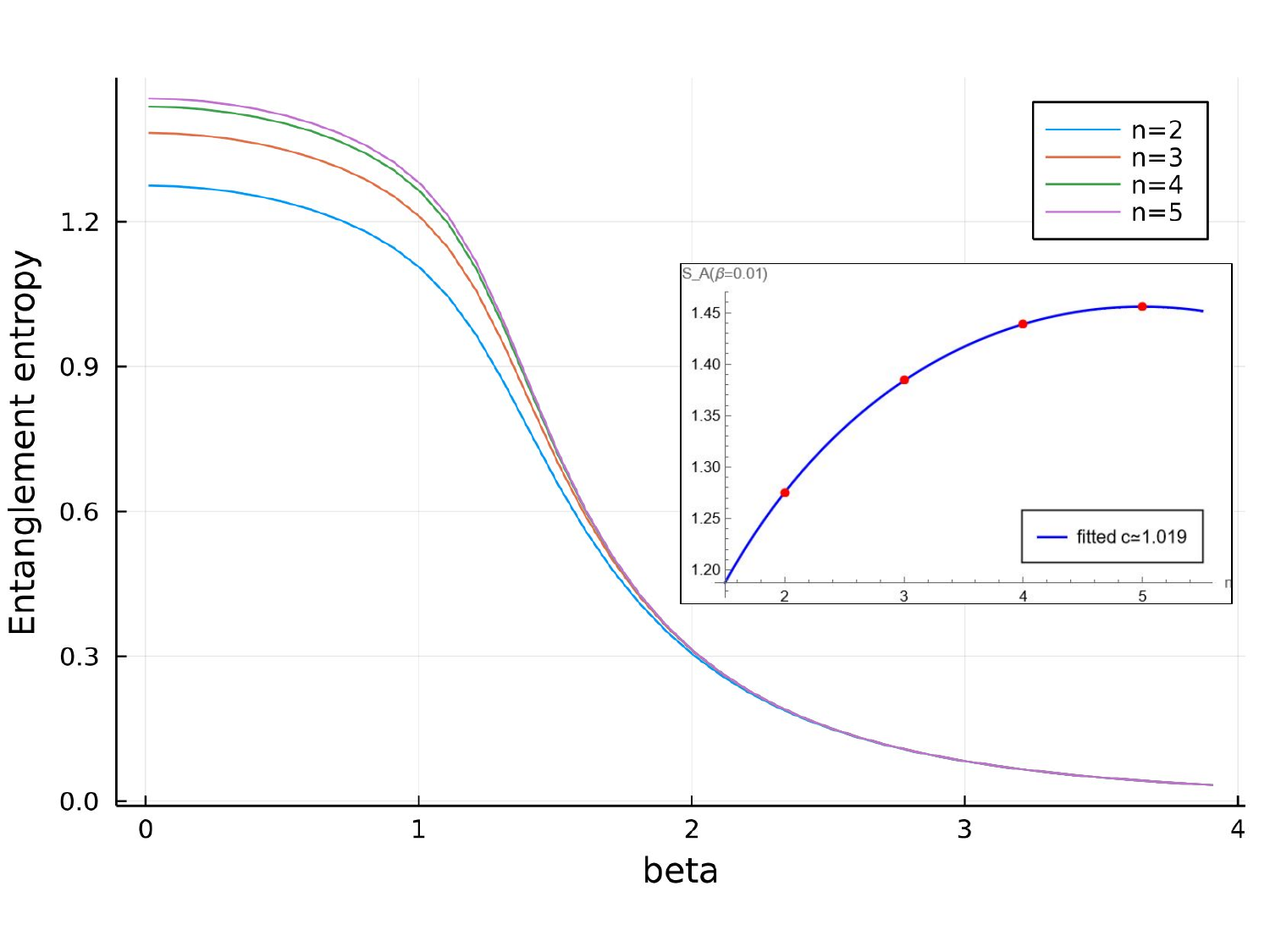}
    }\\
    \subfloat[]{
    \includegraphics[width=.9\linewidth]{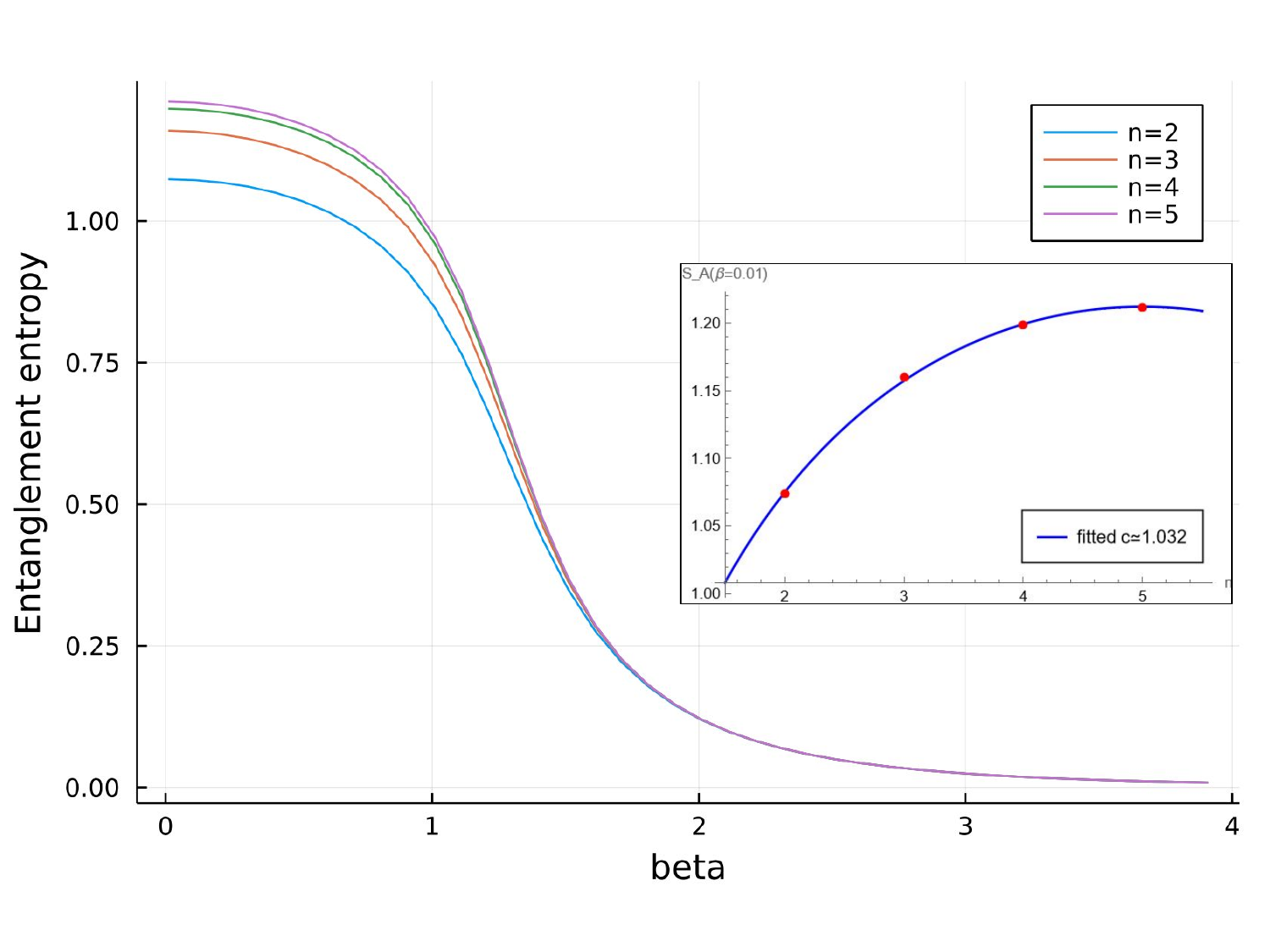}
    }
    \caption{The (a)~von Neumann and (b)~2-Reny entropies for the system with size of $N = 10$ and the subsystem with size of $n = 2, 3, 4, 5$ of the ground state with the periodic boundary conditions.
    The insets are the values of $S_A$ at $\beta = 0.01$ as functions of the subsystem size $n$.
    The value of the central charge $c$ is calculated by fitting the functions Eq.~\eqref{eq:ee-critical-finite-vN} and Eq.~\eqref{eq:ee-critical-finite-Renyi} (blue curves), respectively, giving the estimations of $c \approx 1.019$ for the von Neumann entropies and $c \approx 1.032$ for the 2-Renyi entropies. 
    }
    \label{fig:ee-sg}
\end{figure}

Zhang~\cite{Zhang:2021nhm} computes the von Neumann entropy of the ground state of the Hamiltonian Eq.~\eqref{eq:SG-H} with $h=0$ and open boundary conditions, and it is confirmed that the gapped/gapless transition happens with $L=1$ and the system size of $N= 32, 64, 96, 128$ with the subsystem size $n = N/2$ by means of density matrix renormalization group. 
As we mentioned, the results from~\cite{Zhang:2021dnz, Zhang:2021nhm} indicate that this model with $L=1$ has the transition called infinite-order gaussian transition, which is distinguished from the BKT transition but still closes/opens the mass gap. 
We reproduce this result by computing the von Neumann and 2-Renyi entropies from exact diagonalization of the Hamiltonian with the much smaller system size of $N = 10$ and the sub-system size of $n = 2, 3, 4, 5$ as functions of $\beta$ with two disjoint boundaries (Fig.~\ref{fig:ee-sg}).
In the small $\beta$ region, the entanglement entropies depend on the subsystem size, and the gaps become smaller as $n$ increases, as we expect for the gapless phase. 
On the other hand, in the large $\beta$ region, the entropies become independent of $n$, so the system must be in the gapped phase. 
That we can observe this gapped/gapless transition with only a small system size of $N=10$ means that even with the current or near-future digital quantum device consisting of less than a hundred qubits, we may reproduce interesting physical phenomena related to a continuous field theory. 
Let us note that many efficient quantum algorithms for realizing the ground state on qubits for a given spin Hamiltonian have been proposed such as the variational quantum eigensolver~\cite{Peruzzo2014}, the adiabatic state preparation~\cite{Farhi2000}, and imaginary time evolution~\cite{McArdle2019, Motta2019}, etc, 
as well as that one can efficiently evaluate the 2-Renyi entropy on a digital quantum device by computing the expectation value $\bra{GS}\otimes \bra{GS} SWAP_A \ket{GS}\otimes \ket{GS}$ where $\ket{GS}\otimes \ket{GS}$ is the two copies of the ground state and $SWAP_A$ is the operation that swaps the qubits in the subsystem $A$ between those two copies~\cite{Hastings:2010zka}. 
Such expectation values of a unitary operator can be computed using for example the Hadamard test. 
\\

We can also find the central charge $c$ of this model by simply fitting the values of $S_A$ and $S_A^{(2)}$ in the gapless phase to the functions Eq.~\eqref{eq:ee-critical-finite-vN} and Eq.~\eqref{eq:ee-critical-finite-Renyi}, respectively. 
We fit their values with $\beta = 0.01$, and estimate the value of the central charges as $c \approx 1.019$ for $S_A$ and $c \approx 1.032$ for $S_A^{(2)}$, respectively (the insets of Fig.~\ref{fig:ee-sg}). 
These almost reproduce $c = 1$ which is expected for this class of the model~\cite{Allton1988, DiFrancesco1997}.

\section{\label{sec:Discussion}Discussion}

In this paper, we have discussed a version of the quantum link model with {\em gauged flavor symmetry}~\cite{Schlittgen:2000xg,Bar:2001gz} focusing especially on the problem of a $U(1)$ quantum link.
The main problem we have focused on in this paper is how to realize local degrees of freedom that are effectively bosonic and have a non-trivial symmetry structure realized on them in terms of (fermionic) qubits. 
Generically, these are ingredients that can be used on a variety of field theories, not just for gauge theories. 
Such a choice depends on if one puts the degrees of freedom on a link or on a lattice site, and it would also depend on if one imposes a local symmetry constraint or not, which would involve many links at a time. 
A single link/lattice variable would not know these spatial configurations on its own. 
The main problem with bosonic systems is that they naturally have an infinite-dimensional Hilbert space, even locally. 
This needs to be truncated if it is to be simulated on a quantum computer.

The truncation suggested by the fermionic qubits for $U(1)$ picks a particular quantization of a system that has an $SU(2)$ symmetry on the phase space, but only a $U(1)$ symmetry in the Hamiltonian. 
This gives us a quantum theory on a two-sphere, which is realized by angular momentum operators with a fixed value of $L^2$. 
We showed in the classical theory and in the quantum theory how taking $L^2\to \infty$ results in the Kogut-Susskind Hamiltonian after an appropriate rescaling of the variables. 
The physics is in a compact phase space locally with finite volume in the units of the quantum $\hbar$. 
Taking the volume to infinity can provide the phase space of a tangent bundle on the $U(1)$ manifold if done appropriately. 
We also showed how additional corrections to the Hamiltonian (which can be thought of as higher derivative corrections) could be added so that the naive flux-truncated Kogut-Susskind Hamiltonian for a single variable could be found exactly, rather than approximately at finite cutoff. 
This type of argument suggests that the link variables with gauge flavor symmetry fall in the same universality class as Kogut-Susskind type Hamiltonians do in the appropriate limit, without the need to add these higher-order corrections. 

Generalizing this construction to other non-Abelian symmetry groups seems to require studying the quantization on a complex Grassmannian (a compact phase space) and taking a similar large volume limit in units of $\hbar$.

Some special features were found in the $U(1)$ theory where the original problem with $2M$ fermionic qubits could be reduced to $M$ qubits that are effectively hard bosons: they commute with each other. 
The physics requires that the permutation group between these hard bosons was fully symmetrized between them to faithfully achieve the flavor gauge symmetry. 
The qubit realization of the $U, U^\dagger$ operators resulted in the unique representation that appears from the addition of angular momentum for these variables, with maximal angular momentum.

We studied various versions of the truncated Hamiltonians that differ from each other in the choices that are made for these higher derivative corrections and found good agreement with the quantum rotor (the Kogut-Susskind Hamiltonian with no cutoff). 
They supposedly approximate even for moderate values of the coupling. 
It is interesting to study this property further for other models with non-Abelian symmetry.

We also studied other implementations that are not based  on the fermionic bilinears,  but where the truncation in Hilbert space is done to minimize the number of total qubits, and the embedding is more ad-hoc (there is no natural symmetry action on the qubit degrees of freedom, it needs to be built by hand). 
At least in this sense, one can talk about the efficiency of the implementation in terms of resources.

We applied these ideas to the models with two such $U(1)$ degrees of
freedom as would appear in a chiral $U(1)$ model on a one-dimensional lattice. 
In particular, we showed how a simple truncation could be
implemented in terms of explicit gates on a collection of 12-qubits
(two per site) and showed how the Trotter expansion could be executed
for studying the real-time evolution of a simply prepared initial state.
Furthermore, in the $U(1)$ case, we used exact diagonalization to
argue that the ground state on a lattice of only 10 sites was already
big enough to show non-trivial critical behavior in the entanglement
entropy when varying the coupling constant. 
This suggests that interesting physics at (or near) criticality can be simulated on a modest quantum computer with roughly $\sim 100$ qubits, rather than requiring us to take the large volume limit first.

What is left out in this study is any serious exploration of how a minimal number of qubit per lattice site might preserve the universality class.
Indeed this is a central and very challenging dynamical problem depending on the existence of a second-order critical surface.
While preserving the symplectic algebra on the local field is clearly an attractive requirement, it does not address this problem. As in the classical Ising Hamiltonian with a single qubit per site with $Z_2$ reflection, the collective dynamics across the spatial lattice is sufficient to guarantee universality.

There are many potential routes to universality.
For example, we have also left out the original quantum link conjecture that in an asymptotically free theory (the non-Abelian 2-D  sigma model or the 4-D gauge theory), flavors distributed in an extra dimension are sufficient to guarantee universality.  
Such models break the flavor symmetry but would also reduce the number of quantum gates to be executed to logarithmic
growth in the correlation length. 
Our previous work \cite{Brower:2020huh}, for example, provides such an implementation for a $U(1)$ gauge theory in $2+1$ dimensions. 
One needs to worry that the breaking of the flavor symmetry done in the Hamiltonian does not pollute the infrared physics with new degrees of freedom that are not gapped sufficiently. 
With full gauging of flavor, as we studied here, there are no additional singlet states beyond those required to match the Hilbert space of interest, so the only question is if we approximated the correct Hamiltonian well enough in the low energy sector.
A full treatment of such questions needs to be explored in detail.

\section*{Acknowledgements}
 This work was supported in part by the U.S. Department of Energy (DOE) under Award No. DE-SC0019139 and under  DE-SC0015845 for RCB. 
 HK was supported by Yukawa Institute for Theoretical Physics (YITP) for the expenses during the visit in Summer 2021 and would like to thank their hospitality. 
 HK would like to thank Masazumi Honda and Etsuko Itou for arranging the stay as well as the fruitful discussions during and after the stay. 
HK would also like to thank Suguru Endo, Thomas Lloyd, and Jin Zhang for fruitful discussions and information. 

\bibliographystyle{apsrev4-2}
\bibliography{U1}

\begin{thebibliography}{44}%
\makeatletter
\providecommand \@ifxundefined [1]{%
 \@ifx{#1\undefined}
}%
\providecommand \@ifnum [1]{%
 \ifnum #1\expandafter \@firstoftwo
 \else \expandafter \@secondoftwo
 \fi
}%
\providecommand \@ifx [1]{%
 \ifx #1\expandafter \@firstoftwo
 \else \expandafter \@secondoftwo
 \fi
}%
\providecommand \natexlab [1]{#1}%
\providecommand \enquote  [1]{``#1''}%
\providecommand \bibnamefont  [1]{#1}%
\providecommand \bibfnamefont [1]{#1}%
\providecommand \citenamefont [1]{#1}%
\providecommand \href@noop [0]{\@secondoftwo}%
\providecommand \href [0]{\begingroup \@sanitize@url \@href}%
\providecommand \@href[1]{\@@startlink{#1}\@@href}%
\providecommand \@@href[1]{\endgroup#1\@@endlink}%
\providecommand \@sanitize@url [0]{\catcode `\\12\catcode `\$12\catcode
  `\&12\catcode `\#12\catcode `\^12\catcode `\_12\catcode `\%12\relax}%
\providecommand \@@startlink[1]{}%
\providecommand \@@endlink[0]{}%
\providecommand \url  [0]{\begingroup\@sanitize@url \@url }%
\providecommand \@url [1]{\endgroup\@href {#1}{\urlprefix }}%
\providecommand \urlprefix  [0]{URL }%
\providecommand \Eprint [0]{\href }%
\providecommand \doibase [0]{https://doi.org/}%
\providecommand \selectlanguage [0]{\@gobble}%
\providecommand \bibinfo  [0]{\@secondoftwo}%
\providecommand \bibfield  [0]{\@secondoftwo}%
\providecommand \translation [1]{[#1]}%
\providecommand \BibitemOpen [0]{}%
\providecommand \bibitemStop [0]{}%
\providecommand \bibitemNoStop [0]{.\EOS\space}%
\providecommand \EOS [0]{\spacefactor3000\relax}%
\providecommand \BibitemShut  [1]{\csname bibitem#1\endcsname}%
\let\auto@bib@innerbib\@empty
\bibitem [{\citenamefont {Brambilla}\ \emph {et~al.}(2014)\citenamefont
  {Brambilla} \emph {et~al.}}]{Brambilla:2014jmp}%
  \BibitemOpen
  \bibfield  {author} {\bibinfo {author} {\bibfnamefont {N.}~\bibnamefont
  {Brambilla}} \emph {et~al.},\ }\href
  {https://doi.org/10.1140/epjc/s10052-014-2981-5} {\bibfield  {journal}
  {\bibinfo  {journal} {Eur. Phys. J.}\ }\textbf {\bibinfo {volume} {C74}},\
  \bibinfo {pages} {2981} (\bibinfo {year} {2014})},\ \Eprint
  {https://arxiv.org/abs/1404.3723} {arXiv:1404.3723 [hep-ph]} \BibitemShut
  {NoStop}%
\bibitem [{\citenamefont {Feynman}(1982)}]{feynman1982simulating}%
  \BibitemOpen
  \bibfield  {author} {\bibinfo {author} {\bibfnamefont {R.~P.}\ \bibnamefont
  {Feynman}},\ }\href@noop {} {\bibfield  {journal} {\bibinfo  {journal}
  {International journal of theoretical physics}\ }\textbf {\bibinfo {volume}
  {21}},\ \bibinfo {pages} {467} (\bibinfo {year} {1982})}\BibitemShut
  {NoStop}%
\bibitem [{\citenamefont {Kogut}\ and\ \citenamefont
  {Susskind}(1975)}]{Kogut:1974ag}%
  \BibitemOpen
  \bibfield  {author} {\bibinfo {author} {\bibfnamefont {J.~B.}\ \bibnamefont
  {Kogut}}\ and\ \bibinfo {author} {\bibfnamefont {L.}~\bibnamefont
  {Susskind}},\ }\href {https://doi.org/10.1103/PhysRevD.11.395} {\bibfield
  {journal} {\bibinfo  {journal} {Phys. Rev.}\ }\textbf {\bibinfo {volume}
  {D11}},\ \bibinfo {pages} {395} (\bibinfo {year} {1975})}\BibitemShut
  {NoStop}%
\bibitem [{Note1()}]{Note1}%
  \BibitemOpen
  \bibinfo {note} {On a regular hypercubic lattice, fields on directed links
  $\protect \langle x,y\protect \rangle $ with a positive spatial shift, $y = x
  +\protect \hat \mu $ are often labelled by $U_\mu (x) = U(x,x+\protect \hat
  \mu )$ and the somewhat awkward backward link by $U^\protect \dag _\mu (x, x
  -\protect \hat \mu ) = U(x+\protect \hat \mu ,x)$. The discrete curl for the
  magnetic term is then the path ordered product on each square: $U_{\mu \nu
  }(x) = U(x,x +\mu ) U(x+\mu ,x + \nu ) U(x+\nu ,x+\mu ) U(x+\nu
  ,x)$.}\BibitemShut {Stop}%
\bibitem [{\citenamefont {Chandrasekharan}\ and\ \citenamefont
  {Wiese}(1997)}]{Chandrasekharan:1996ih}%
  \BibitemOpen
  \bibfield  {author} {\bibinfo {author} {\bibfnamefont {S.}~\bibnamefont
  {Chandrasekharan}}\ and\ \bibinfo {author} {\bibfnamefont {U.~J.}\
  \bibnamefont {Wiese}},\ }\href
  {https://doi.org/10.1016/S0550-3213(97)00006-0} {\bibfield  {journal}
  {\bibinfo  {journal} {Nucl. Phys. B}\ }\textbf {\bibinfo {volume} {492}},\
  \bibinfo {pages} {455} (\bibinfo {year} {1997})},\ \Eprint
  {https://arxiv.org/abs/hep-lat/9609042} {arXiv:hep-lat/9609042} \BibitemShut
  {NoStop}%
\bibitem [{\citenamefont {Brower}\ \emph {et~al.}(1999)\citenamefont {Brower},
  \citenamefont {Chandrasekharan},\ and\ \citenamefont
  {Wiese}}]{Brower:1997ha}%
  \BibitemOpen
  \bibfield  {author} {\bibinfo {author} {\bibfnamefont {R.}~\bibnamefont
  {Brower}}, \bibinfo {author} {\bibfnamefont {S.}~\bibnamefont
  {Chandrasekharan}},\ and\ \bibinfo {author} {\bibfnamefont {U.~J.}\
  \bibnamefont {Wiese}},\ }\href {https://doi.org/10.1103/PhysRevD.60.094502}
  {\bibfield  {journal} {\bibinfo  {journal} {Phys. Rev.}\ }\textbf {\bibinfo
  {volume} {D60}},\ \bibinfo {pages} {094502} (\bibinfo {year} {1999})},\
  \Eprint {https://arxiv.org/abs/hep-th/9704106} {arXiv:hep-th/9704106
  [hep-th]} \BibitemShut {NoStop}%
\bibitem [{\citenamefont {Beard}\ \emph {et~al.}(1998)\citenamefont {Beard},
  \citenamefont {Brower}, \citenamefont {Chandrasekharan}, \citenamefont
  {Chen}, \citenamefont {Tsapalis},\ and\ \citenamefont
  {Wiese}}]{Beard:1997ic}%
  \BibitemOpen
  \bibfield  {author} {\bibinfo {author} {\bibfnamefont {B.~B.}\ \bibnamefont
  {Beard}}, \bibinfo {author} {\bibfnamefont {R.~C.}\ \bibnamefont {Brower}},
  \bibinfo {author} {\bibfnamefont {S.}~\bibnamefont {Chandrasekharan}},
  \bibinfo {author} {\bibfnamefont {D.}~\bibnamefont {Chen}}, \bibinfo {author}
  {\bibfnamefont {A.}~\bibnamefont {Tsapalis}},\ and\ \bibinfo {author}
  {\bibfnamefont {U.~J.}\ \bibnamefont {Wiese}},\ }\bibfield  {booktitle}
  {\emph {\bibinfo {booktitle} {{Contents of LAT97 proceedings}}},\ }\href
  {https://doi.org/10.1016/S0920-5632(97)00900-6} {\bibfield  {journal}
  {\bibinfo  {journal} {Nucl. Phys. Proc. Suppl.}\ }\textbf {\bibinfo {volume}
  {63}},\ \bibinfo {pages} {775} (\bibinfo {year} {1998})},\ \Eprint
  {https://arxiv.org/abs/hep-lat/9709120} {arXiv:hep-lat/9709120 [hep-lat]}
  \BibitemShut {NoStop}%
\bibitem [{\citenamefont {Brower}\ \emph {et~al.}(2004)\citenamefont {Brower},
  \citenamefont {Chandrasekharan}, \citenamefont {Riederer},\ and\
  \citenamefont {Wiese}}]{Brower:2003vy}%
  \BibitemOpen
  \bibfield  {author} {\bibinfo {author} {\bibfnamefont {R.}~\bibnamefont
  {Brower}}, \bibinfo {author} {\bibfnamefont {S.}~\bibnamefont
  {Chandrasekharan}}, \bibinfo {author} {\bibfnamefont {S.}~\bibnamefont
  {Riederer}},\ and\ \bibinfo {author} {\bibfnamefont {U.~J.}\ \bibnamefont
  {Wiese}},\ }\href {https://doi.org/10.1016/j.nuclphysb.2004.06.007}
  {\bibfield  {journal} {\bibinfo  {journal} {Nucl. Phys.}\ }\textbf {\bibinfo
  {volume} {B693}},\ \bibinfo {pages} {149} (\bibinfo {year} {2004})},\ \Eprint
  {https://arxiv.org/abs/hep-lat/0309182} {arXiv:hep-lat/0309182 [hep-lat]}
  \BibitemShut {NoStop}%
\bibitem [{\citenamefont {Bravyi}\ and\ \citenamefont
  {Kitaev}(2002)}]{Bravyi_2002}%
  \BibitemOpen
  \bibfield  {author} {\bibinfo {author} {\bibfnamefont {S.~B.}\ \bibnamefont
  {Bravyi}}\ and\ \bibinfo {author} {\bibfnamefont {A.~Y.}\ \bibnamefont
  {Kitaev}},\ }\href {https://doi.org/10.1006/aphy.2002.6254} {\bibfield
  {journal} {\bibinfo  {journal} {Annals of Physics}\ }\textbf {\bibinfo
  {volume} {298}},\ \bibinfo {pages} {210–226} (\bibinfo {year}
  {2002})}\BibitemShut {NoStop}%
\bibitem [{Note2()}]{Note2}%
  \BibitemOpen
  \bibinfo {note} {Here we mean that we exactly preserve the local symmetry of
  gauge transformations on a local link,}\BibitemShut {NoStop}%
\bibitem [{\citenamefont {Wiese}(2021)}]{wiese2021quantum}%
  \BibitemOpen
  \bibfield  {author} {\bibinfo {author} {\bibfnamefont {U.-J.}\ \bibnamefont
  {Wiese}},\ }\href@noop {} {\bibinfo {title} {From quantum link models to
  d-theory: A resource efficient framework for the quantum simulation and
  computation of gauge theories}} (\bibinfo {year} {2021}),\ \Eprint
  {https://arxiv.org/abs/2107.09335} {arXiv:2107.09335 [hep-lat]} \BibitemShut
  {NoStop}%
\bibitem [{\citenamefont {Liu}\ and\ \citenamefont
  {Chandrasekharan}(2021)}]{https://doi.org/10.48550/arxiv.2112.02090}%
  \BibitemOpen
  \bibfield  {author} {\bibinfo {author} {\bibfnamefont {H.}~\bibnamefont
  {Liu}}\ and\ \bibinfo {author} {\bibfnamefont {S.}~\bibnamefont
  {Chandrasekharan}},\ }\href {https://doi.org/10.48550/ARXIV.2112.02090}
  {\bibinfo {title} {Qubit regularization and qubit embedding algebras}}
  (\bibinfo {year} {2021})\BibitemShut {NoStop}%
\bibitem [{\citenamefont {Schlittgen}\ and\ \citenamefont
  {Wiese}(2001)}]{Schlittgen:2000xg}%
  \BibitemOpen
  \bibfield  {author} {\bibinfo {author} {\bibfnamefont {B.}~\bibnamefont
  {Schlittgen}}\ and\ \bibinfo {author} {\bibfnamefont {U.~J.}\ \bibnamefont
  {Wiese}},\ }\href {https://doi.org/10.1103/PhysRevD.63.085007} {\bibfield
  {journal} {\bibinfo  {journal} {Phys. Rev.}\ }\textbf {\bibinfo {volume}
  {D63}},\ \bibinfo {pages} {085007} (\bibinfo {year} {2001})},\ \Eprint
  {https://arxiv.org/abs/hep-lat/0012014} {arXiv:hep-lat/0012014 [hep-lat]}
  \BibitemShut {NoStop}%
\bibitem [{\citenamefont {Bhattacharya}\ \emph {et~al.}(2021)\citenamefont
  {Bhattacharya}, \citenamefont {Buser}, \citenamefont {Chandrasekharan},
  \citenamefont {Gupta},\ and\ \citenamefont {Singh}}]{Bhattacharya:2020gpm}%
  \BibitemOpen
  \bibfield  {author} {\bibinfo {author} {\bibfnamefont {T.}~\bibnamefont
  {Bhattacharya}}, \bibinfo {author} {\bibfnamefont {A.~J.}\ \bibnamefont
  {Buser}}, \bibinfo {author} {\bibfnamefont {S.}~\bibnamefont
  {Chandrasekharan}}, \bibinfo {author} {\bibfnamefont {R.}~\bibnamefont
  {Gupta}},\ and\ \bibinfo {author} {\bibfnamefont {H.}~\bibnamefont {Singh}},\
  }\href {https://doi.org/10.1103/PhysRevLett.126.172001} {\bibfield  {journal}
  {\bibinfo  {journal} {Phys. Rev. Lett.}\ }\textbf {\bibinfo {volume} {126}},\
  \bibinfo {pages} {172001} (\bibinfo {year} {2021})},\ \Eprint
  {https://arxiv.org/abs/2012.02153} {arXiv:2012.02153 [hep-lat]} \BibitemShut
  {NoStop}%
\bibitem [{\citenamefont {Zhang}\ \emph {et~al.}(2021)\citenamefont {Zhang},
  \citenamefont {Meurice},\ and\ \citenamefont {Tsai}}]{Zhang:2021dnz}%
  \BibitemOpen
  \bibfield  {author} {\bibinfo {author} {\bibfnamefont {J.}~\bibnamefont
  {Zhang}}, \bibinfo {author} {\bibfnamefont {Y.}~\bibnamefont {Meurice}},\
  and\ \bibinfo {author} {\bibfnamefont {S.~W.}\ \bibnamefont {Tsai}},\ }\href
  {https://doi.org/10.1103/PhysRevB.103.245137} {\bibfield  {journal} {\bibinfo
   {journal} {Phys. Rev.}\ }\textbf {\bibinfo {volume} {B103}},\ \bibinfo
  {pages} {245137} (\bibinfo {year} {2021})}\BibitemShut {NoStop}%
\bibitem [{\citenamefont {Bar}\ \emph {et~al.}(2002)\citenamefont {Bar},
  \citenamefont {Brower}, \citenamefont {Schlittgen},\ and\ \citenamefont
  {Wiese}}]{Bar:2001gz}%
  \BibitemOpen
  \bibfield  {author} {\bibinfo {author} {\bibfnamefont {O.}~\bibnamefont
  {Bar}}, \bibinfo {author} {\bibfnamefont {R.}~\bibnamefont {Brower}},
  \bibinfo {author} {\bibfnamefont {B.}~\bibnamefont {Schlittgen}},\ and\
  \bibinfo {author} {\bibfnamefont {U.~J.}\ \bibnamefont {Wiese}},\ }\bibfield
  {booktitle} {\emph {\bibinfo {booktitle} {{Lattice field theory. Proceedings,
  19th International Symposium, Lattice 2001, Berlin, Germany, August 19-24,
  2001}}},\ }\href {https://doi.org/10.1016/S0920-5632(01)01916-8} {\bibfield
  {journal} {\bibinfo  {journal} {Nucl. Phys. Proc. Suppl.}\ }\textbf {\bibinfo
  {volume} {106}},\ \bibinfo {pages} {1019} (\bibinfo {year} {2002})},\ \Eprint
  {https://arxiv.org/abs/hep-lat/0110148} {arXiv:hep-lat/0110148 [hep-lat]}
  \BibitemShut {NoStop}%
\bibitem [{\citenamefont {Raynal}\ \emph {et~al.}(2010)\citenamefont {Raynal},
  \citenamefont {Kalev}, \citenamefont {Suzuki},\ and\ \citenamefont
  {Englert}}]{Raynal_Kalev_Suzuki_Englert_2010}%
  \BibitemOpen
  \bibfield  {author} {\bibinfo {author} {\bibfnamefont {P.}~\bibnamefont
  {Raynal}}, \bibinfo {author} {\bibfnamefont {A.}~\bibnamefont {Kalev}},
  \bibinfo {author} {\bibfnamefont {J.}~\bibnamefont {Suzuki}},\ and\ \bibinfo
  {author} {\bibfnamefont {B.-G.}\ \bibnamefont {Englert}},\ }\bibfield
  {journal} {\bibinfo  {journal} {Physical Review A}\ }\textbf {\bibinfo
  {volume} {81}},\ \href {https://doi.org/10.1103/physreva.81.052327}
  {10.1103/physreva.81.052327} (\bibinfo {year} {2010})\BibitemShut {NoStop}%
\bibitem [{\citenamefont {Albert}\ \emph {et~al.}(2017)\citenamefont {Albert},
  \citenamefont {Pascazio},\ and\ \citenamefont {Devoret}}]{Albert:2017}%
  \BibitemOpen
  \bibfield  {author} {\bibinfo {author} {\bibfnamefont {V.~V.}\ \bibnamefont
  {Albert}}, \bibinfo {author} {\bibfnamefont {S.}~\bibnamefont {Pascazio}},\
  and\ \bibinfo {author} {\bibfnamefont {M.~H.}\ \bibnamefont {Devoret}},\
  }\href {https://doi.org/10.1088/1751-8121/aa9314} {\bibfield  {journal}
  {\bibinfo  {journal} {Journal of Physics A: Mathematical and Theoretical}\
  }\textbf {\bibinfo {volume} {50}},\ \bibinfo {pages} {504002} (\bibinfo
  {year} {2017})}\BibitemShut {NoStop}%
\bibitem [{\citenamefont {Berenstein}(2005)}]{Berenstein:2004hw}%
  \BibitemOpen
  \bibfield  {author} {\bibinfo {author} {\bibfnamefont {D.}~\bibnamefont
  {Berenstein}},\ }\href {https://doi.org/10.1103/PhysRevD.71.085001}
  {\bibfield  {journal} {\bibinfo  {journal} {Phys. Rev.}\ }\textbf {\bibinfo
  {volume} {D71}},\ \bibinfo {pages} {085001} (\bibinfo {year} {2005})},\
  \Eprint {https://arxiv.org/abs/hep-th/0409115} {arXiv:hep-th/0409115
  [hep-th]} \BibitemShut {NoStop}%
\bibitem [{\citenamefont {Berenstein}\ and\ \citenamefont
  {de~Mello~Koch}(2019)}]{Berenstein:2019esh}%
  \BibitemOpen
  \bibfield  {author} {\bibinfo {author} {\bibfnamefont {D.}~\bibnamefont
  {Berenstein}}\ and\ \bibinfo {author} {\bibfnamefont {R.}~\bibnamefont
  {de~Mello~Koch}},\ }\href {https://doi.org/10.1007/JHEP03(2019)185}
  {\bibfield  {journal} {\bibinfo  {journal} {JHEP}\ }\textbf {\bibinfo
  {volume} {03}},\ \bibinfo {pages} {185}},\ \Eprint
  {https://arxiv.org/abs/1903.01628} {arXiv:1903.01628 [hep-th]} \BibitemShut
  {NoStop}%
\bibitem [{\citenamefont {Gross}\ and\ \citenamefont
  {Taylor}(1993)}]{Gross:1993hu}%
  \BibitemOpen
  \bibfield  {author} {\bibinfo {author} {\bibfnamefont {D.~J.}\ \bibnamefont
  {Gross}}\ and\ \bibinfo {author} {\bibfnamefont {W.}~\bibnamefont {Taylor}},\
  }\href {https://doi.org/10.1016/0550-3213(93)90403-C} {\bibfield  {journal}
  {\bibinfo  {journal} {Nucl. Phys.}\ }\textbf {\bibinfo {volume} {B400}},\
  \bibinfo {pages} {181} (\bibinfo {year} {1993})},\ \Eprint
  {https://arxiv.org/abs/hep-th/9301068} {arXiv:hep-th/9301068 [hep-th]}
  \BibitemShut {NoStop}%
\bibitem [{Note3()}]{Note3}%
  \BibitemOpen
  \bibinfo {note} {Also the two-torus.}\BibitemShut {Stop}%
\bibitem [{Note4()}]{Note4}%
  \BibitemOpen
  \bibinfo {note} {The Casimir of the charge state is $\ell ^2$, so this is
  also a cutoff on the maximum value of the Casimir.}\BibitemShut {Stop}%
\bibitem [{Note5()}]{Note5}%
  \BibitemOpen
  \bibinfo {note} {As a side note, the clock model is the quantization one gets
  by using a two-torus cylinder geometry, rather than the sphere. The algebra
  is then very similar to that of a fuzzy torus, but with this choice of $E$
  there is a discontinuity of the classical variable at the gluing of the upper
  and lower part of the cylinder that avoids a doubling of low energy degrees
  of freedom.}\BibitemShut {Stop}%
\bibitem [{\citenamefont {Creutz}\ \emph {et~al.}(1983)\citenamefont {Creutz},
  \citenamefont {Jacobs},\ and\ \citenamefont {Rebbi}}]{CREUTZ1983201}%
  \BibitemOpen
  \bibfield  {author} {\bibinfo {author} {\bibfnamefont {M.}~\bibnamefont
  {Creutz}}, \bibinfo {author} {\bibfnamefont {L.}~\bibnamefont {Jacobs}},\
  and\ \bibinfo {author} {\bibfnamefont {C.}~\bibnamefont {Rebbi}},\ }\href
  {https://doi.org/https://doi.org/10.1016/0370-1573(83)90016-9} {\bibfield
  {journal} {\bibinfo  {journal} {Physics Reports}\ }\textbf {\bibinfo {volume}
  {95}},\ \bibinfo {pages} {201} (\bibinfo {year} {1983})}\BibitemShut
  {NoStop}%
\bibitem [{Note6()}]{Note6}%
  \BibitemOpen
  \bibinfo {note} {In the usual spin notation, we have that $m=j_z+j$, where
  $j$ is the spin of the representation and $j_z$ is the $z$ eigenvalue of
  angular momentum.}\BibitemShut {Stop}%
\bibitem [{Note7()}]{Note7}%
  \BibitemOpen
  \bibinfo {note} {The exponential formats still enjoy the quantum advantage
  once we turn to the theory with $d > 0$ spatial dimensions}\BibitemShut
  {NoStop}%
\bibitem [{\citenamefont {Takahashi}(2009)}]{Takahashi_2009}%
  \BibitemOpen
  \bibfield  {author} {\bibinfo {author} {\bibfnamefont {Y.}~\bibnamefont
  {Takahashi}},\ }\href {https://doi.org/10.1587/transfun.E92.A.1276}
  {\bibfield  {journal} {\bibinfo  {journal} {IEICE Transactions}\ }\textbf
  {\bibinfo {volume} {92-A}},\ \bibinfo {pages} {1276} (\bibinfo {year}
  {2009})}\BibitemShut {NoStop}%
\bibitem [{\citenamefont {Takahashi}\ and\ \citenamefont
  {Kunihiro}(2008)}]{Takahashi_2008}%
  \BibitemOpen
  \bibfield  {author} {\bibinfo {author} {\bibfnamefont {Y.}~\bibnamefont
  {Takahashi}}\ and\ \bibinfo {author} {\bibfnamefont {N.}~\bibnamefont
  {Kunihiro}},\ }\href@noop {} {\bibfield  {journal} {\bibinfo  {journal}
  {Quantum Info. Comput.}\ }\textbf {\bibinfo {volume} {8}},\ \bibinfo {pages}
  {636–649} (\bibinfo {year} {2008})}\BibitemShut {NoStop}%
\bibitem [{\citenamefont {Smilga}(2001)}]{Smilga:2001ck}%
  \BibitemOpen
  \bibfield  {author} {\bibinfo {author} {\bibfnamefont {A.~V.}\ \bibnamefont
  {Smilga}},\ }\href {https://doi.org/10.1142/4443} {\emph {\bibinfo {title}
  {{Lectures on quantum chromodynamics}}}}\ (\bibinfo  {publisher} {WSP},\
  \bibinfo {address} {Singapore},\ \bibinfo {year} {2001})\BibitemShut
  {NoStop}%
\bibitem [{\citenamefont {Coleman}(1975)}]{Coleman1975}%
  \BibitemOpen
  \bibfield  {author} {\bibinfo {author} {\bibfnamefont {S.}~\bibnamefont
  {Coleman}},\ }\href {https://doi.org/10.1103/PhysRevD.11.2088} {\bibfield
  {journal} {\bibinfo  {journal} {Phys. Rev. D}\ }\textbf {\bibinfo {volume}
  {11}},\ \bibinfo {pages} {2088} (\bibinfo {year} {1975})}\BibitemShut
  {NoStop}%
\bibitem [{\citenamefont {Mishra}\ \emph {et~al.}(2019)\citenamefont {Mishra},
  \citenamefont {Thompson}, \citenamefont {Pooser},\ and\ \citenamefont
  {Siopsis}}]{Mishra:2019xbh}%
  \BibitemOpen
  \bibfield  {author} {\bibinfo {author} {\bibfnamefont {C.}~\bibnamefont
  {Mishra}}, \bibinfo {author} {\bibfnamefont {S.}~\bibnamefont {Thompson}},
  \bibinfo {author} {\bibfnamefont {R.}~\bibnamefont {Pooser}},\ and\ \bibinfo
  {author} {\bibfnamefont {G.}~\bibnamefont {Siopsis}}\ }\href
  {https://doi.org/10.1088/2058-9565/ab8f63} {10.1088/2058-9565/ab8f63}
  (\bibinfo {year} {2019}),\ \Eprint {https://arxiv.org/abs/1912.07767}
  {arXiv:1912.07767 [quant-ph]} \BibitemShut {NoStop}%
\bibitem [{\citenamefont {Gogolin}\ \emph {et~al.}(2004)\citenamefont
  {Gogolin}, \citenamefont {Nersesian},\ and\ \citenamefont
  {Tsvelik}}]{Gogolin:2004rp}%
  \BibitemOpen
  \bibfield  {author} {\bibinfo {author} {\bibfnamefont {A.~O.}\ \bibnamefont
  {Gogolin}}, \bibinfo {author} {\bibfnamefont {A.~A.}\ \bibnamefont
  {Nersesian}},\ and\ \bibinfo {author} {\bibfnamefont {A.~M.}\ \bibnamefont
  {Tsvelik}},\ }\href@noop {} {\emph {\bibinfo {title} {{Bosonization and
  strongly correlated systems}}}}\ (\bibinfo {year} {2004})\BibitemShut
  {NoStop}%
\bibitem [{\citenamefont {Hastings}(2007)}]{Hastings:2007}%
  \BibitemOpen
  \bibfield  {author} {\bibinfo {author} {\bibfnamefont {M.~B.}\ \bibnamefont
  {Hastings}},\ }\href {https://doi.org/10.1088/1742-5468/2007/08/p08024}
  {\bibfield  {journal} {\bibinfo  {journal} {Journal of Statistical Mechanics:
  Theory and Experiment}\ }\textbf {\bibinfo {volume} {2007}},\ \bibinfo
  {pages} {P08024} (\bibinfo {year} {2007})}\BibitemShut {NoStop}%
\bibitem [{\citenamefont {Calabrese}\ and\ \citenamefont
  {Cardy}(2004)}]{Calabrese:2004eu}%
  \BibitemOpen
  \bibfield  {author} {\bibinfo {author} {\bibfnamefont {P.}~\bibnamefont
  {Calabrese}}\ and\ \bibinfo {author} {\bibfnamefont {J.~L.}\ \bibnamefont
  {Cardy}},\ }\href {https://doi.org/10.1088/1742-5468/2004/06/P06002}
  {\bibfield  {journal} {\bibinfo  {journal} {J. Stat. Mech.}\ }\textbf
  {\bibinfo {volume} {0406}},\ \bibinfo {pages} {P06002} (\bibinfo {year}
  {2004})},\ \Eprint {https://arxiv.org/abs/hep-th/0405152}
  {arXiv:hep-th/0405152} \BibitemShut {NoStop}%
\bibitem [{\citenamefont {Zhang}(2021)}]{Zhang:2021nhm}%
  \BibitemOpen
  \bibfield  {author} {\bibinfo {author} {\bibfnamefont {J.}~\bibnamefont
  {Zhang}},\ }\href@noop {} {\  (\bibinfo {year} {2021})},\ \Eprint
  {https://arxiv.org/abs/2108.09966} {arXiv:2108.09966 [quant-ph]} \BibitemShut
  {NoStop}%
\bibitem [{\citenamefont {Peruzzo}\ \emph {et~al.}(2014)\citenamefont
  {Peruzzo}, \citenamefont {McClean}, \citenamefont {Shadbolt}, \citenamefont
  {Yung}, \citenamefont {Zhou}, \citenamefont {Love}, \citenamefont
  {Aspuru-Guzik},\ and\ \citenamefont {O’Brien}}]{Peruzzo2014}%
  \BibitemOpen
  \bibfield  {author} {\bibinfo {author} {\bibfnamefont {A.}~\bibnamefont
  {Peruzzo}}, \bibinfo {author} {\bibfnamefont {J.}~\bibnamefont {McClean}},
  \bibinfo {author} {\bibfnamefont {P.}~\bibnamefont {Shadbolt}}, \bibinfo
  {author} {\bibfnamefont {M.-H.}\ \bibnamefont {Yung}}, \bibinfo {author}
  {\bibfnamefont {X.-Q.}\ \bibnamefont {Zhou}}, \bibinfo {author}
  {\bibfnamefont {P.~J.}\ \bibnamefont {Love}}, \bibinfo {author}
  {\bibfnamefont {A.}~\bibnamefont {Aspuru-Guzik}},\ and\ \bibinfo {author}
  {\bibfnamefont {J.~L.}\ \bibnamefont {O’Brien}},\ }\bibfield  {journal}
  {\bibinfo  {journal} {Nature Communications}\ }\textbf {\bibinfo {volume}
  {5}},\ \href {https://doi.org/10.1038/ncomms5213} {10.1038/ncomms5213}
  (\bibinfo {year} {2014})\BibitemShut {NoStop}%
\bibitem [{\citenamefont {Farhi}\ \emph {et~al.}(2000)\citenamefont {Farhi},
  \citenamefont {Goldstone}, \citenamefont {Gutmann},\ and\ \citenamefont
  {Sipser}}]{Farhi2000}%
  \BibitemOpen
  \bibfield  {author} {\bibinfo {author} {\bibfnamefont {E.}~\bibnamefont
  {Farhi}}, \bibinfo {author} {\bibfnamefont {J.}~\bibnamefont {Goldstone}},
  \bibinfo {author} {\bibfnamefont {S.}~\bibnamefont {Gutmann}},\ and\ \bibinfo
  {author} {\bibfnamefont {M.}~\bibnamefont {Sipser}},\ }\href@noop {}
  {\bibfield  {journal} {\bibinfo  {journal} {arXiv preprint quant-ph/0001106}\
  } (\bibinfo {year} {2000})}\BibitemShut {NoStop}%
\bibitem [{\citenamefont {McArdle}\ \emph {et~al.}(2019)\citenamefont
  {McArdle}, \citenamefont {Jones}, \citenamefont {Endo}, \citenamefont {Li},
  \citenamefont {Benjamin},\ and\ \citenamefont {Yuan}}]{McArdle2019}%
  \BibitemOpen
  \bibfield  {author} {\bibinfo {author} {\bibfnamefont {S.}~\bibnamefont
  {McArdle}}, \bibinfo {author} {\bibfnamefont {T.}~\bibnamefont {Jones}},
  \bibinfo {author} {\bibfnamefont {S.}~\bibnamefont {Endo}}, \bibinfo {author}
  {\bibfnamefont {Y.}~\bibnamefont {Li}}, \bibinfo {author} {\bibfnamefont
  {S.~C.}\ \bibnamefont {Benjamin}},\ and\ \bibinfo {author} {\bibfnamefont
  {X.}~\bibnamefont {Yuan}},\ }\bibfield  {journal} {\bibinfo  {journal} {npj
  Quantum Information}\ }\textbf {\bibinfo {volume} {5}},\ \href
  {https://doi.org/10.1038/s41534-019-0187-2} {10.1038/s41534-019-0187-2}
  (\bibinfo {year} {2019})\BibitemShut {NoStop}%
\bibitem [{\citenamefont {Motta}\ \emph {et~al.}(2019)\citenamefont {Motta},
  \citenamefont {Sun}, \citenamefont {Tan}, \citenamefont {O’Rourke},
  \citenamefont {Ye}, \citenamefont {Minnich}, \citenamefont {Brandão},\ and\
  \citenamefont {Chan}}]{Motta2019}%
  \BibitemOpen
  \bibfield  {author} {\bibinfo {author} {\bibfnamefont {M.}~\bibnamefont
  {Motta}}, \bibinfo {author} {\bibfnamefont {C.}~\bibnamefont {Sun}}, \bibinfo
  {author} {\bibfnamefont {A.~T.~K.}\ \bibnamefont {Tan}}, \bibinfo {author}
  {\bibfnamefont {M.~J.}\ \bibnamefont {O’Rourke}}, \bibinfo {author}
  {\bibfnamefont {E.}~\bibnamefont {Ye}}, \bibinfo {author} {\bibfnamefont
  {A.~J.}\ \bibnamefont {Minnich}}, \bibinfo {author} {\bibfnamefont {F.~G.
  S.~L.}\ \bibnamefont {Brandão}},\ and\ \bibinfo {author} {\bibfnamefont
  {G.~K.-L.}\ \bibnamefont {Chan}},\ }\href
  {https://doi.org/10.1038/s41567-019-0704-4} {\bibfield  {journal} {\bibinfo
  {journal} {Nature Physics}\ }\textbf {\bibinfo {volume} {16}},\ \bibinfo
  {pages} {205–210} (\bibinfo {year} {2019})}\BibitemShut {NoStop}%
\bibitem [{\citenamefont {Hastings}\ \emph {et~al.}(2010)\citenamefont
  {Hastings}, \citenamefont {Gonz\'alez}, \citenamefont {Kallin},\ and\
  \citenamefont {Melko}}]{Hastings:2010zka}%
  \BibitemOpen
  \bibfield  {author} {\bibinfo {author} {\bibfnamefont {M.~B.}\ \bibnamefont
  {Hastings}}, \bibinfo {author} {\bibfnamefont {I.}~\bibnamefont
  {Gonz\'alez}}, \bibinfo {author} {\bibfnamefont {A.~B.}\ \bibnamefont
  {Kallin}},\ and\ \bibinfo {author} {\bibfnamefont {R.~G.}\ \bibnamefont
  {Melko}},\ }\href {https://doi.org/10.1103/PhysRevLett.104.157201} {\bibfield
   {journal} {\bibinfo  {journal} {Phys. Rev. Lett.}\ }\textbf {\bibinfo
  {volume} {104}},\ \bibinfo {pages} {157201} (\bibinfo {year} {2010})},\
  \Eprint {https://arxiv.org/abs/1001.2335} {arXiv:1001.2335 [cond-mat.str-el]}
  \BibitemShut {NoStop}%
\bibitem [{\citenamefont {Allton}\ and\ \citenamefont
  {Hamer}(1988)}]{Allton1988}%
  \BibitemOpen
  \bibfield  {author} {\bibinfo {author} {\bibfnamefont {C.~R.}\ \bibnamefont
  {Allton}}\ and\ \bibinfo {author} {\bibfnamefont {C.~J.}\ \bibnamefont
  {Hamer}},\ }\href {https://doi.org/10.1088/0305-4470/21/10/019} {\bibfield
  {journal} {\bibinfo  {journal} {Journal of Physics A: Mathematical and
  General}\ }\textbf {\bibinfo {volume} {21}},\ \bibinfo {pages} {2417}
  (\bibinfo {year} {1988})}\BibitemShut {NoStop}%
\bibitem [{\citenamefont {Di~Francesco}\ \emph {et~al.}(1997)\citenamefont
  {Di~Francesco}, \citenamefont {Mathieu},\ and\ \citenamefont
  {S{\'e}n{\'e}chal}}]{DiFrancesco1997}%
  \BibitemOpen
  \bibfield  {author} {\bibinfo {author} {\bibfnamefont {P.}~\bibnamefont
  {Di~Francesco}}, \bibinfo {author} {\bibfnamefont {P.}~\bibnamefont
  {Mathieu}},\ and\ \bibinfo {author} {\bibfnamefont {D.}~\bibnamefont
  {S{\'e}n{\'e}chal}},\ }\bibinfo {title} {Conformal field theory}\ (\bibinfo
  {publisher} {Springer New York},\ \bibinfo {address} {New York, NY},\
  \bibinfo {year} {1997})\BibitemShut {NoStop}%
\bibitem [{\citenamefont {Brower}\ \emph {et~al.}(2020)\citenamefont {Brower},
  \citenamefont {Berenstein},\ and\ \citenamefont {Kawai}}]{Brower:2020huh}%
  \BibitemOpen
  \bibfield  {author} {\bibinfo {author} {\bibfnamefont {R.~C.}\ \bibnamefont
  {Brower}}, \bibinfo {author} {\bibfnamefont {D.}~\bibnamefont {Berenstein}},\
  and\ \bibinfo {author} {\bibfnamefont {H.}~\bibnamefont {Kawai}},\ }\href
  {https://doi.org/10.22323/1.363.0112} {\bibfield  {journal} {\bibinfo
  {journal} {PoS}\ }\textbf {\bibinfo {volume} {LATTICE2019}},\ \bibinfo
  {pages} {112} (\bibinfo {year} {2020})},\ \Eprint
  {https://arxiv.org/abs/2002.10028} {arXiv:2002.10028 [hep-lat]} \BibitemShut
  {NoStop}%
\end{thebibliography}%

\end{document}